\documentclass{aa}
\usepackage[varg]{txfonts}
\usepackage{hyperref}
\usepackage{units}
\usepackage{amssymb}
\usepackage[english]{babel}
\usepackage{color}
\usepackage{array}
\usepackage{booktabs}
\usepackage{graphicx}
\usepackage{array}
\usepackage{subfig}
\usepackage{fnpos}
\usepackage{listings}
\usepackage{booktabs}

\usepackage[T1]{fontenc}
\hypersetup{
    colorlinks=true,
    linkcolor=blue,
    filecolor=magenta,      
    urlcolor=blue,
}

\newcolumntype{M}[1]{>{\centering\arraybackslash}m{#1}}
\newcolumntype{N}{@{}m{0pt}@{}}

\newcommand{\HI}{H\textsc{i}}

\newcommand{\kms}{km\,s$^{-1}$}
\newcommand{\snint}{S/N$_{\mathrm{int}}$}

\begin{document}

\title{On the completeness and reliability of visual source extraction : an examination of eight thousand data cubes by eye}

\author{Rhys Taylor\inst{\ref{inst1}}}

\institute{Astronomical Institute of the Czech Academy of Sciences, Bo\v{c}ní II 1401/1, 141 00, Prague, Czech Republic \\ \email{rhysyt@gmail.com}\label{inst1}}



\abstract
{Source extraction in \HI{} radio surveys is still often performed using visual, by-eye inspection, but the efficacy of such procedures lacks rigorous quantitative assessment due to the laborious nature of the procedure. Algorithmic methods are often preferred due to their repeatable results and speed.}
{To quantitatively assess the completeness and reliability of visual source extraction by using a suitably large sample of artificial sources and a comparatively rapid source extraction tool, and compare the results with those from automatic techniques.}
{A dedicated source extraction tool was modified to significantly reduce the cataloguing speed. I injected 4,232 sources into a total of 8,500 emission-free data cubes, with at most one source per cube. Sources covered a wide range of signal-to-noise and velocity widths. I blindly searched all cubes for the sources, measuring the completeness and reliability for pairs of signal-to-noise and line width values. Smaller control tests were performed to account for the possible biases in the search, which gave results in good agreement with the main experiment. I also searched cubes injected with artificial sources using algorithmic extractors, and compare these results with a set of catalogues independently reported from real observational data which were searched with different automatic methods.}
{I find that the results of visual extraction follow a tight relation between integrated signal-to-noise and completeness. Visual extraction compares favourably in efficacy with the algorithmic methods, tending to recover a higher fraction of fainter sources.}
{Visual source extraction can be a surprisingly rapid procedure which gives higher completeness levels than automatic techniques, giving predictable, quantifiable results which are not strongly subject to the whims of the observer. For recovering the faintest features, algorithmic extractors can be competitive with visual inspection but cannot yet out-perform it, though their advantage in speed can be a significant compensating factor.}


\maketitle

\section{Introduction}
\label{sec:intro}
All source catalogues can be described in terms of their \textit{completeness} and \textit{reliability}. Completeness describes what fraction of sources present in an inspected data set are recovered by the source extraction procedure. Reliability is the fraction of the sources in the catalogue which are real, astrophysical objects. Understanding both of these is essential for interpreting any survey. An incomplete survey may well be biased, potentially only detecting highly atypical objects. Conversely an unreliable catalogue will contain many objects which are not even real. Both completeness and reliability need to be maximised to ensure the most robust scientific interpretations of the data.

Completeness and reliability are in principle independent parameters; a change in source extraction method that affects one parameter will not necessarily affect the other. In practice the situation is more subtle. For example, to generate a catalogue which is 100\% complete is trivial : one simply declares that all pixels in a data set correspond to a source. However this obviously comes with a severe penalty in terms of reliability, which might in that case approach 0\% if few real sources happen to be present. Vice-versa, by deciding that only the brightest pixels correspond to real sources, one may well ensure very high reliability levels but suffer from extremely low completeness, potentially missing most of the recoverable sources (or at least the most interesting ones, which have an uncanny tendency to also be the faintest).

In extragalactic \HI{} data sets it has long been common to produce catalogues using visual source extraction, for example \cite{anotherdeephi}, \cite{cenahi}, \cite{hideep}, \cite{alfazoa} and \cite{parkeszoa}. Many others use a combination of procedures, e.g. automatic techniques which are either supplemented with visual extraction or use visual inspection as a verification step -- for example, \cite{adbs} (who note a greater reliability of visual extraction), \cite{hipassmain}, \cite{hipassnorth}, and \cite{auds}. The advantage of visual extraction is that it is extremely simple to implement : one simply pans through a data set (be that a data cube or individual spectra) looking for whatever appears likely, by some criteria, to correspond to a source, and recording the coordinates whenever something is found. The disadvantages are that it can be extremely laborious and time-consuming, and is supposedly highly subjective : what may appear as a source to one person may not be noticed by another, and even the same person may not necessarily notice the same source on different inspections of the same data set. Minimising this subjectivity is obviously highly desirable.

Despite the importance of quantifying completeness and reliability, relatively few attempts have been made to quantify how (un)successful visual source extraction can be. One suitable way to do this is to inject a large number of artificial sources into a data set and visually search for them. This controlled method means that one knows exactly how many sources are present without any contamination from real objects, which would themselves suffer from unknown levels of completeness and reliability in the source extraction procedure. The difficulty has been that the number of sources required for this procedure makes it a daunting prospect. There have however been some estimates. \cite{adbs} inject artificial sources into real data cubes which also contain real signals. They use a combination of recovery rates of the artificial signals and follow-up observations of the potential real sources to estimate completeness and reliability, noting the difficulty in determining an accurate completeness function. In \cite{agesvc2} we compared the number of sources found by visual procedures and a selection of algorithms, while \cite{lotsoffakes} injected 400 artificial sources to measure the recovery rates; \cite{saint} (hereafter S07) use a similar procedure for testing an automatic source extractor. 

When analysing AGES (Arecibo Galaxy Environment Survey - \citealt{ages}) data, we have used a variety of source extraction techniques over the years. In our earliest studies (\citealt{ages}, \citealt{agesii}, \citealt{agesiii}), we would visually inspect the data using \textit{kvis} (\citealt{kvis}). Two or three people would examine each data cube, all looking slice-by-slice along all three projections of the cube. Source coordinates were recorded by hand and the resulting tables cross-matched, with the final catalogue list assigning source status (accepted, rejected, or requiring follow-up observations) based on the number of people who found the source and in how many projections. In subsequent papers we have generally reduced this to a single person performing visual extraction in combination with a variety of automatic algorithms, especially GLADoS (Galaxy Line Analysis for Detection of Sources; \citealt{agesvc1}) and \textsc{duchamp} (\citealt{duchamp}).

With the advent of FRELLED (FITS Realtime Explorer of Low Latency in Every Dimension; \citealt{agesvii, T15, FRELLED5}) visual extraction was greatly simplified. Since the user can interactively both mask and catalogue sources, each object takes only a few seconds to define. This was shown to make the visual extraction process about 50 times faster than the old method. With a specially-modified version of FRELLED configured for handling a large set of data cubes, this makes it practical to attempt to search for a number of objects large enough to accurately determine the completeness and reliability of visual extraction. Although automatic techniques such as the SoFiA program (Source Finding Application; \citealt{SOFIA}) have become increasingly popular (and will surely continue to do so given the ever-larger surveys such as \citealt{WALLABY} and \citealt{fasthi}), visual extraction remains useful. While  major statistical results might be obtained through purely algorithmic methods, it is difficult to believe that individually interesting objects will ever be reported without at least some visual analytic assessment (see \citealt{visualsarenice}).

For typical AGES and WAVES (Widefield Arecibo Virgo Environment Survey; \citealt{waves}) data sets, a full catalogue from visual extraction with FRELLED takes about a day to produce. This is so fast that there is almost no point in \textit{not} undertaking a by-eye search, which has the substantial benefit of giving enormous familiarity with the data set. It also remains a valuable teaching tool both for students and new observers to understand their data sets.

Our source \textit{verification} methodology has also varied. The gold standard is follow-up observations, employed throughout all AGES studies where possible. For practical reasons this is mainly used only to verify the existence of uncertain sources, and only occasionally to obtain more accurate measurements of their properties (e.g. total mass, line width). Our reliability for these sources, after follow-up, has varied from 28\%-85\%. Our ideal is around 50\% as this is a good balance between searching for more risky detections and not wasting telescope time. Sources at the extremes of (un)reliability are generally judged so based on their S/N (signal-to-noise ratio, see sections \ref{sec:setup} and \ref{sec:comprelvar}), quality of the data in which they reside (e.g. presence or otherwise of radio frequency interference - RFI), inspection of the individual polarisations which comprise the data (see in particular \citealt{agesvc1}) and correspondence with multi-wavelength data. 

Here I describe an experiment to quantify the completeness and reliability of visual source extraction, briefly referred to in \cite{AGESLeo} and \cite{boris} but here described in full. By injecting (approximately) 4,000 sources into over 8,000 small data cubes, I performed a blind search across a wide range of line widths and peak fluxes. The remainder of this work describes this investigation. In section \ref{sec:setup} I describe the modifications necessary to FRELLED and the details of the search procedure used. Section \ref{sec:results} presents the results of the search, quantifying the completeness and reliability as a function of different parameters. Tests for biases in the main result are described in section \ref{sec:biasproof}. Using the results of the main experiment, section \ref{sec:gocompare} compares the performance of visual and selected automatic source extraction algorithms. Finally section \ref{sec:sum} summarises the results and discusses the implications regarding AGES and other surveys. I argue that visual source extraction is a powerful and rapid technique which is not as subjective as may be supposed.

\section{Experimental setup}
\label{sec:setup}
The typical process of using FRELLED for visual source extraction is described in detail in \cite{agesvii} and \cite{T15}; see also \cite {FRELLED5} for the latest version. In brief, FRELLED allows the user to click on the location of a source and create a virtual object. This can be manipulated in terms of scale and position to mask the source and its coordinates extracted for further analysis. FRELLED can display the data in 3D volumetrically, which makes it easy to define the mask object in 3D space, but also in 2D as sequences of images. The latter is better for sensitivity, but since the source mask must be checked in every slice in which the source is believed to be present, it can be slower than the 3D mode. However, here the aim is not to accurately measure the source parameters, but purely an exercise in detectability : to quantify completeness and reliability. This meant that the data was inspected exclusively in 2D mode. All the user had to do was click on the position of a potential source, restricting the operations to the absolute minimum in order to maximise speed. Being based around AGES data in which the vast majority of detections are \textit{at most} marginally spatially resolved, the setup here uses purely point sources. The code and data sets used for the experiments described throughout this work are given in the `data availability' section.

Much of the particulars of this approach were arrived at through trial and error. One option would have been to use a single large data cube full of sources, which would be a close mimic to real source extraction. However injecting enough sources to span the necessary range and fidelity of S/N and line width proved difficult. This would require the user to mask the sources reasonably accurately, which adds considerably to the time (and effort !) required. 

The final strategy adopted was to use many small data cubes in which sources of different line widths and S/N were injected. Note that here S/N refers to peak S/N value, i.e. peak flux / \textit{rms}, and this should be assumed to be the case throughout this work -- other formulations of S/N are used but given a different notation (see section \ref{sec:comprelvar}). For each combination of line width and S/N, 100 empty data cubes were used, and either one or zero sources of these parameters were injected into each. Whether a source was injected or not was determined at random. Alternatively exactly 50 cubes could have been randomly selected but it was thought that this would give the user too much information about the probability of whether a cube would contain a source or not. Source position was randomised in each cube with the only restriction that it should be contained entirely within the data. 

\subsection{Artificial data} 
\label{ref:data}
To be certain of recovering only the sources injected and avoid contamination from real sources, two options are available : either to use purely artificial data, or to use a region of data in which we have high confidence that no real galaxies are present. The first may be too idealised since real data is seldom perfectly Gaussian (but see \citealt{SKADC} who model noise with more sophisticated methods); various artifacts are often present which require different processes to remove (see \citealt{fading}). The latter is difficult to obtain. Fortunately, we do have one observational data set with a large volume in which any real \HI{} emission is extremely unlikely. Using the Mock spectrometer in the NGC 7448 field (\citealt{agesiv, agesvii}), the full 300 MHz bandwidth available to the ALFA (Arecibo L-band Feed Array) instrument was observed, spanning the velocity range -23,500 to +49,200 \kms{}.

Detections are present in the higher-redshift data which will be presented in a future paper. In contrast the blueshifted data extends to velocities so low that the chances of any \HI{} emission is practically nil. For this experiment I extracted the data from -3,000 to -6,000\,\kms{}. While OH megamaser emission can occur in this range, with the OH line shifted to the corresponding frequency if at a velocity of about 47,000\,\kms{}, this is sufficiently rare that this too can be neglected. \cite{masers} reported the discovery of 5 OH megamasers in 2,800 square degrees of the ALFALFA survey, (Arecibo Legacy Fast ALFA; \citealt{ALFALFA}) meaning the we should not expect to detect a single such object in the entire AGES survey of 200 square degrees.

In this subset, each spectrum was fitted by a second-order polynomial which was then removed, a standard AGES procedure as described in \cite{agesvii}. Each pixel was also divided by the \textit{rms} of its spectrum, effectively recasting the cube from flux to S/N values. The data is everywhere approximately Gaussian but with a higher \textit{rms} at the edges owing to the arrangement of the ALFA multi-beam receiver and the AGES drift-scan survey strategy. This means that converting the data to S/N (a technique also described in \citealt{agesvii}), has the advantage of ensuring that the data values span the same range in every pixel. More importantly, using S/N values is beneficial here as they can be easily converted to flux values depending on the survey sensitivity and source distance. 

\begin{figure}[t]
\begin{center}
\includegraphics[width=90mm]{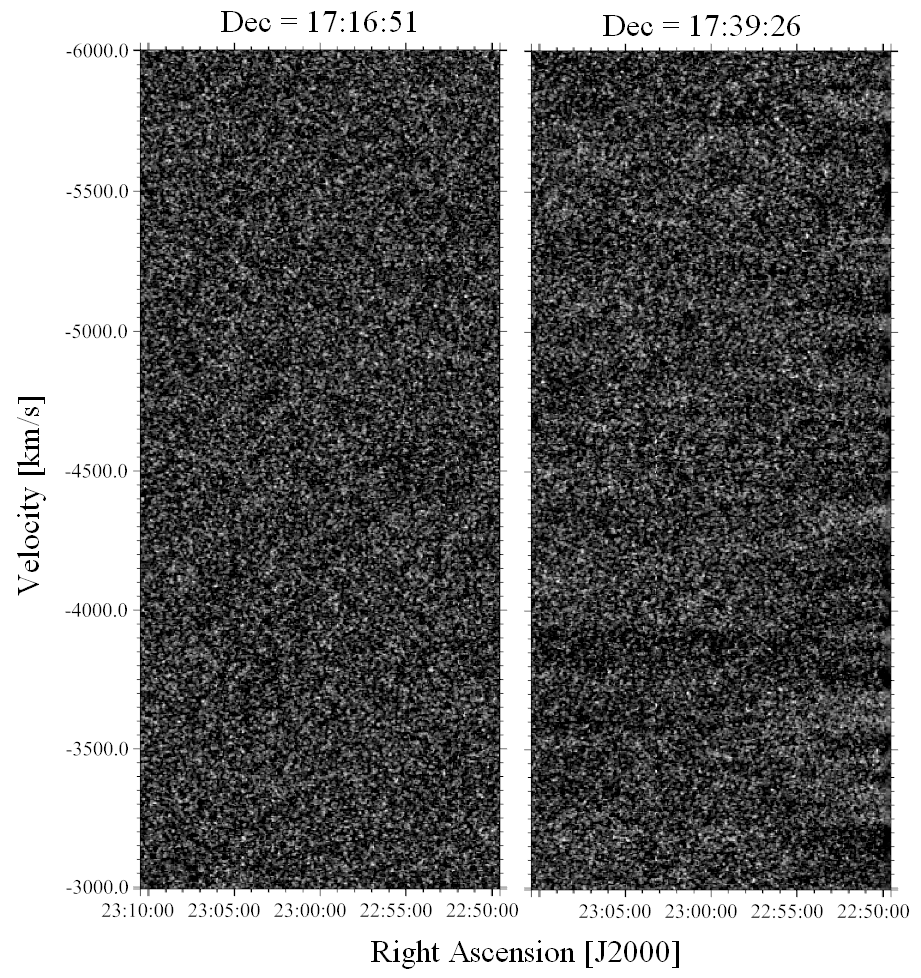}
\caption[ex]{Example PV-diagrams from the master cube after the processing described in section \ref{ref:data}, with the data display range being a S/N range of -1.0 to 5.0 as described in section \ref{sec:specialfrelled}. Two different declination slices are labelled. The left was chosen as an example of the typical data quality, where the noise is uniform. The right shows a more atypical region in which structures in the noise are clearly visible, a problem which affects about 5\% of the data.}
\label{fig:masterdata}
\end{center}
\end{figure}

Two example position-velocity (PV) diagrams from this cube are shown in figure \ref{fig:masterdata}. As discussed in \cite{fading}, the bandpass subtraction used here does not greatly affect the \textit{rms} of the data, but \textit{rms} is not a perfect measure of data quality : it does not depend at all on \textit{coherency} of structures present within the data\footnote{An example of this is shown in figure 2 in \citealt{fading}, which shows a dramatic difference in visual appearance of the data after cleaning but only a modest improvement in the \textit{rms}. This effect can be understood with a simple thought experiment : if one were to re-arrange all the pixels in a data cube in flux order, it is obvious that distinct structures would be visible in the data despite the \textit{rms} not having changed at all.}. While most of the data resembles the left panel of figure \ref{fig:masterdata}, with uniform and homogenous noise, about 5\% of the data by volume shows the coherent structures apparent in the right panel. This is fairly typical of AGES data in which (for a myriad of reasons) things occasionally go wrong with a small number of scans, e.g. RFI, faulty instruments, etc. These imperfections are however rather weak, visible at less than 5$\sigma$ (see section \ref{sec:specialfrelled}) and cover a broad range in R.A. -- it is doubtful they could be mistaken for galaxies. This data set is a good representation of AGES data in general, with some regions having unfortunately structured noise, but only at a low level and occupying a small fractional volume of the survey regions.

The master empty cube has dimensions 300$\times$100$\times$700 pixels (\textit{x}, \textit{y}, \textit{z} axes equivalent to R.A., Declination, velocity). From this, random subsets were extracted of dimensions 100$\times$20$\times$100 pixels. The y-axis (corresponding to declination) was chosen to be large enough that the user would have to pan through the different slices of data along this axis, but small enough that this could be done quickly. Although this setup does have some restrictions (see section \ref{sec:numberbias}), it forms a realistic approximation of the search for individual sources; using multiple cubes per pair of source parameters effectively reproduces the search in a large data cube but without the need to mask each object to avoid confusion.  

The injected sources consisted of 2D Gaussian profiles of FWHM of 3.5 pixels, reflecting the standard AGES unresolved sources and data parameters (a 3.5\arcmin{} Arecibo beam with the data gridded into 1\arcmin{} pixels; for automated extraction which includes resolved sources, see \citealt{SKADC}, some limited visual experiments are given in \citealt{fading}). Each source profile spanned 10$\times$10 spatial pixels, large enough (given their brightness) that the emission at the edges contributed negligibly to the data values -- thus giving a smooth and seamless transition between the real and artificial data. Sources were given a simple top hat spectral profile of the required line width (for the choice of using a top hat profile as opposed to a more realistic shape, see section \ref{sec:comprelvar}), with the AGES data gridded as standard into channels of 5\,\kms{} width. 

Subset cubes were extracted from the master empty cube after the baseline subtraction had been applied but before the standard Hanning smoothing to 10\,\kms{} resolution. This was because it was found that if the sources were injected with sharp spectral profile edges, they were immediately much more obvious than real sources, even when their line width and total flux were very low. Applying the Hanning smoothing afterwards, on the individual subset cubes after injecting the sources, was shown to eliminate this effect, so that the injected sources `blended in' with the data and strongly resembled real objects.

The velocity width of the sources ranged from 25 to 350\,\kms{} in steps of 25\,\kms{}. The S/N ranged from 1.5 to 6.0 in steps of 0.25. Theoretically this would lead to 266 pairs of values corresponding to 26,600 data cubes with approximately 13,000 sources injected, but fortunately in practice it proved unnecessary to fully explore this frankly outlandish parameter space. For any pair of values, if one increases either line width or S/N, both completeness and reliability increase essentially monotonically. When one approaches 100\% for both completeness and reliability, there is no point in continuing because making the sources either larger or brighter cannot really change anything beyond that. I therefore stopped whenever the resulting completeness and reliability both exceeded about 95\%, beyond this the required effort would bring little gain. A few exceptions were made to this general rule, such as the the S/N values above 4.0 at line widths of 50\,\kms{} -- these were tested due to the much lower completeness/reliability (C/R) values in the adjacent column of width 25\,\kms{}.

The lower limit on the velocity width was chosen to reflect real sources. With the \HI{} line itself largely resulting from emission at 10,000\,K, the chance of finding galaxies with emission below 10\,\kms{} line width is negligible. In the full ALFALFA catalogue of 31,502 objects (\citealt{AA100}), only 73 (0.2\%) have widths (W50) lower than 20\,\kms{} (2,165 have widths lower than 50\,\kms{}, 7\%). While it would be possible to explore even lower S/N values, at the lowest combination of S/N and line width values the completeness and reliability obtained were both 0\%, and I believe the results from the explored parameter space are already sufficient to draw conclusions.

\subsection{Source extraction method} 
\label{sec:specialfrelled}
For this experiment FRELLED was modified to reduce its functionality to the bare minimum. For each set of data cubes, the conversion into a FRELLED-compatible format was done ahead of actually opening the files, which greatly reduces the loading time. A further gain in speed was obtained by only using the XZ projection (equivalent to RA and velocity) as this has been found through experience to be the most useful way to inspect the data : in XY (sky position) sources only show up as points, whereas in velocity they appear much more elongated. The displayed data range values were fixed to a minimum of -1.0 and maximum of 5.0 in all cases (using a linear greyscale colour scheme for maximum contrast). These are values typical of those used in searching real AGES data cubes when looking for faint emission, based on experience : using a lower maximum data value would make the fainter sources a bit brighter but at the significant cost of making noise appear brighter as well, while a higher maximum data value would simply make the fainter sources harder to see without enhancing the visibility of the brighter ones.

It is not likely that this display range could be further optimised to have any significant beneficial affects on the results here (and there would obviously be no point in exploring a range designed to give worse outcomes !). I note, though, that while we seldom change the display range during our main visual cataloguing phase, we frequently alter it when assessing individual sources. This is more of use in analysing source properties rather than cataloguing them, e.g. to search for extensions between and from galaxies. The search method used here is therefore a good approximation to the main cataloguing procedure used by AGES, but does not capture the full range of inspection techniques used in analysis.

A simple bash script was created to sequentially open all FRELLED files for each set of subset cubes. On opening the file, the user could pan through the data slices as normal, but when they thought they had found a source, the procedure differed from the usual source extraction procedure. Normally the user would place a mask region and adjust its scaling. Here instead they only placed the 3D cursor, either placing it at the position of a source if they believed one to be present, or outside the data if they believed the data cube contained no source. Having made their decision, when ready they would press a button which would either load and display a mask region highlighting the source or inform them that no source was present, as appropriate. This would also record the result, accepting that the user had found the source provided they placed the cursor anywhere within the region used for injecting the source. They would then close FRELLED and the next file would automatically open. 

\begin{figure*}[t]
\begin{center}
\includegraphics[width=180mm]{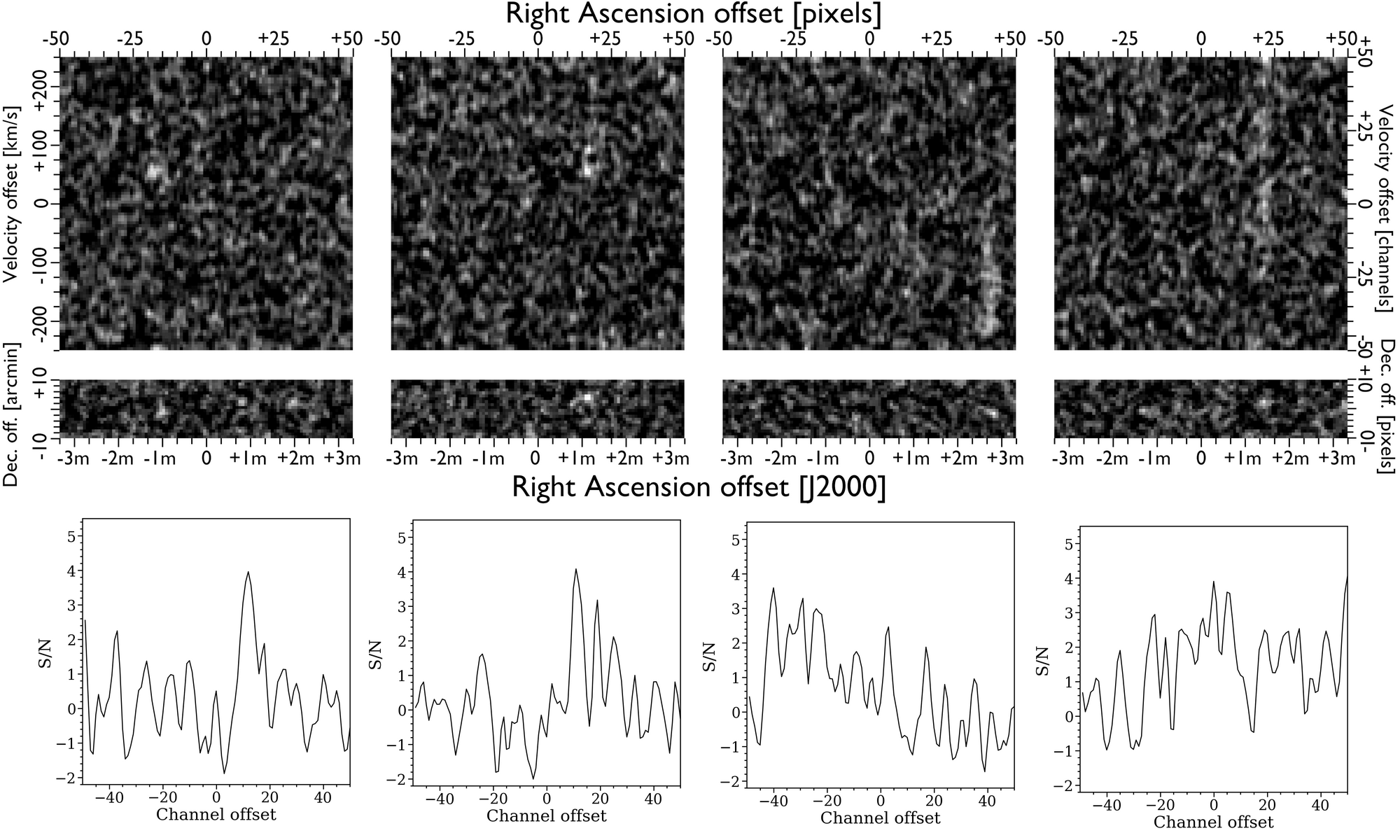}
\caption[ex]{Examples of artificial sources injected into real, empty data cubes. The top row shows the PV diagrams, the middle row shows the sky projections, and the bottom row shows the spectra at the location of the source. World coordinates are given relative to the centre of each data set. For both the PV diagrams and sky projections, the data slices are chosen to show where the sources are most clearly visible (in most velocity channels they are considerably fainter, usually indistinguishable from the noise). From left to right : line widths are 50, 100, 200 and 350\,\kms{}; peak S/N values are 1.75, 1.75, 1.75 and 1.50. All of these sources were detected in the search. Sources are deliberately not marked and locating them is left as a exercise for the reader.}
\label{fig:examples}
\end{center}
\end{figure*}

Some examples of the artificial sources in their data cubes are shown in figure \ref{fig:examples}. Although not used in the search, this includes for comparison the sky projections and spectra. For the broader sources the relatively short spectral baselines are clearly a disadvantage, making the sources less clearly visible in the spectra than they would be in a more typical data set. These wide, faint sources are more obvious to the eye when viewing the FITS data directly, especially in the PV diagrams rather than the sky projections.

This precise method and parameter space was chosen after much experimentation. Batches of 100 data cubes were found to be fast enough that each set of values could be inspected in a reasonably quick time (typically 20-25 minutes for visual inspection of each set of 100 cubes, which required an additional 3 minutes for file setup) which allowed for breaks as necessary; for an inspection of the number of data cubes considered here, the need to avoid visual fatigue becomes important. Showing the user the source (if present) after they had made their decision does give them more information about what they're looking for than might be ideal (in effect training the eye as to what to search for), however in a real data set there are almost at least some bright sources which act as reference. Without any of these it becomes too easy to be lost in the visual fog - one needs \textit{some} clue as to what one is looking for or the experience is unendurably frustrating\footnote{Being a visual search process, one needs to take much more account of the psychological, emotional experience than in using a pure algorithm. However as I will show, once this is done the results can be rigorously objective.}. Moreover, as will be shown in section 3, when source parameters fall below a certain threshold, completeness drops rapidly : no amount of visual training can make the truly invisible visible. The impact of having the user see their results for each source is therefore likely to be small.

On completing the inspection of a pair of values, the choice of which pair of data values to inspect next was generally to increase or decrease either the line width or S/N value by one incremental step (25\,\kms{} or 0.25 respectively). This is for the same reason as above, so that the user remains aware of the basic appearance of what they're searching for. Switching to completely random pairs of values would lose all reference, though it would more accurately reflect the real source extraction procedure. At least some degree of randomness was included however : the initial pair of values was chosen randomly, and at times the switch to different sequences was made arbitrarily.

While the various sources of potential bias (foreknowledge of the source characteristics and their limited number per cube, similarity of sources between different searches, the small size of the cubes to search) are believed to have a low impact for the reasons discussed, concern as to the realism of this experiment nevertheless remains legitimate. Several control tests are described in section \ref{sec:biasproof} which account for these potential problems both individually and collectively.

\section{Results}
\label{sec:results}
The results of the experiment are shown in table \ref{tab:bigtable}. The total number of data cubes searched was 8,500, with a total of 4,232 injected sources. The number injected per S/N-width pair ranged from 35 to 63, with the number of claimed detections per pair (that is, how many sources I thought I had spotted) ranging from 19 to 60. The maximum number of false positives (spurious detections) was 36 and the maximum number of false negatives (sources present but not detected) was 38. 

At the extremes of low S/N and line width, the table does not quite probe the full range of C/R values : it would be necessary to increase the line widths at the lowest S/N values to reach 100/100, and conversely at the lowest line width it would be necessary to increase the S/N to reach 100/100. Importantly, however, the whole table does span the full range of C/R values, from 0/0 to 100/100.

\begin{table*}[t]
\tiny
\begin{center}
\caption[bigtable]{Completeness / reliability values for each set of sources of given parameters.}
\label{tab:bigtable}
\addtolength{\tabcolsep}{-0.2em}
\begin{tabular}{c c c c c c c c c c c c c c c c}\\
\toprule	
\textsc{Velocity width} & 25 & 50 & 75 & 100 & 125 & 150 & 175 & 200 & 225 & 250 & 275 & 300 & 325 & 350\\
\textsc{[\kms{}, n]} & 5 & 10 & 15 & 20 & 25 & 30 & 35 & 40 & 45 & 50 & 55 & 60 & 65 & 70\\
\toprule
\textsc{S/N ratio} & & & & & & & & & & & & & & &\\  
 1.50 & 0/0   & 4/7   &	21/29 &	27/30 &	51/46  & 42/49 & 67/77 & 73/82 & 82/76 & 85/83 & 94/78 & 94/98 & 100/98 & 98/100\\	
 1.75 & 4/5   & 10/19 & 42/38 & 51/44 & 56/78  & 74/64 & 76/94 & 91/93 & 88/95 & 94/98 & 96/98 & 100/100 & -/- & -/-\\
 2.00 & 14/22 & 26/30 & 39/62 & 70/80 & 80/88  & 87/93 & 96/96 & 96/94 & 98/98 & -/-   & -/-   & -/- & -/- & -/-\\ 
 2.25 & 10/15 &	38/45 &	68/72 &	80/88 &	89/98  & 98/98 & 100/96 & 98/100 & -/- & -/-   & -/-   & -/- & -/- & -/-\\
 2.50 & 15/37 &	44/63 &	78/89 &	85/100 & 95/91 & 98/100	& -/-   & -/-   & -/-  & -/-   & -/-   & -/- & -/- & -/-\\
 2.75 & 16/21 &	53/74 &	82/95 &	88/93 &	98/100 & 100/100 & -/-  & -/-   & -/-  & -/-   & -/-   & -/- & -/- & -/-\\
 3.00 &	18/36 & 71/87 &	84/91 &	91/100 & 100/100 & 100/100 & -/- & -/-  & -/-  & -/-   & -/-   & -/- & -/- & -/-\\
 3.25 &	22/40 &	78/84 &	98/98 &	100/98 & -/-     & -/-    & -/-  & -/-  & -/-  & -/-   & -/-   & -/- & -/- & -/-\\
 3.50 &	29/62 &	88/94 &	100/98 & -/-   & -/-     & -/-    & -/-  & -/-  & -/-  & -/-   & -/-   & -/- & -/- & -/-\\
 3.75 &	28/54 &	100/95 & -/-   & -/-   & -/-     & -/-    & -/-  & -/-  & -/-  & -/-   & -/-   & -/- & -/- & -/-\\
 4.00 &	51/67 &	98/91 &	100/100 & -/-  &	-/-  & -/-    & -/-   & -/- & -/-  & -/-   & -/-   & -/- & -/- & -/-\\
 4.25 & 69/69 & 100/100 & -/- & -/- & -/- & -/- & -/- & -/- & -/- & -/- & -/- & -/- & -/- & -/- \\
 4.50 & 73/79 & 100/98 & -/- & -/- & -/- & -/- & -/- & -/- & -/- & -/- & -/- & -/- & -/- & -/- \\
 4.75 & 80/90 & -/- & -/- & -/- & -/- & -/- & -/- & -/- & -/- & -/- & -/- & -/- & -/- & -/- \\
 5.00 & 79/88 & 100/98 & -/- & -/- & -/- & -/- & -/- & -/- & -/- & -/- & -/- & -/- & -/- & -/- \\
 5.25 & 90/88 & -/- & -/- & -/- & -/- & -/- & -/- & -/- & -/- & -/- & -/- & -/- & -/- & -/- \\
 5.50 & 88/88 & -/- & -/- & -/- & -/- & -/- & -/- & -/- & -/- & -/- & -/- & -/- & -/- & -/- \\
 5.75 & 98/94 & -/- & -/- & -/- & -/- & -/- & -/- & -/- & -/- & -/- & -/- & -/- & -/- & -/- \\
 6.00 & 96/94 & 100/100 & -/- & -/- & -/- & -/- & -/- & -/- & -/- & -/- & -/- & -/- & -/- & -/- \\
\end{tabular}
\tablefoot{Entries of `-/-' were not examined, but are assumed to be approximately 100/100 based on adjacent entries of lower S/N and/or line widths (these are not shown in any of the subsequent plots, which only use actual measured values). Velocity widths are given in units of both \kms{} (upper row) and number of channels (lower row).}
\end{center}
\end{table*}	

\subsection{Statistical limitations}
\label{sec:numberbias}
In a real survey there is no constraint on how many sources may be present (save astrophysical limits) and the user has no foreknowledge of what may be present (save a general familiarity with the surveys). The process here is a good approximation to the search for individual sources : at each point in a data cube, one must make the binary choice as to whether a source is present or not, just as in each single data cube in this experiment, one makes the same choice. The difference is that in a real cube one may decide an \textit{unlimited} number of sources are present whereas here one knows ahead of time, if only approximately, the total number of sources, and one can never guess that there are more sources present than the number of data cubes (100 per value pair). In addition, each source found automatically affects both completeness and reliability, meaning they are not independent. This is true in a real survey as well, but here the finite number of sources places much greater restrictions on how completeness can scale with reliability.

\begin{figure}[t]
\begin{center}
\includegraphics[width=90mm]{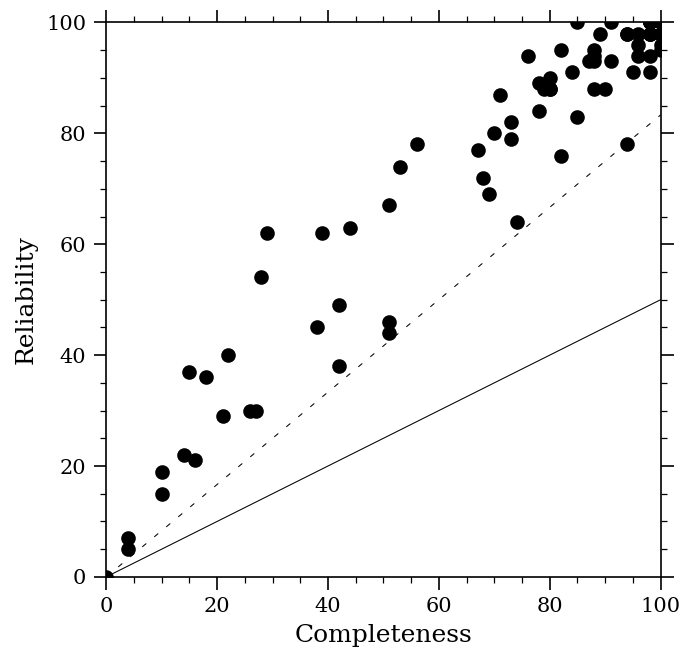}
\caption[c-r]{Reliability as a function of completeness. A clear correlation is seen, but this is partly due to numerical limits. The solid line shows the minimum reliability if a spurious source had been found in every data cube in addition to the actual sources found, which would be always 50\% of the completeness. The dashed line shows a more realistic estimate, accounting for the fact that no more than 60 detections were ever made.}
\label{fig:c-r}
\end{center}
\end{figure}

Figure \ref{fig:c-r} attempts to assess the impact of this, plotting all values of completeness and reliability for each pair of S/N and line width values. A clear correlation is evident, however this is largely a statistical artifact. Suppose, for instance, that in some pair of values exactly 50 sources were injected into the 100 data cubes. In a real survey there would be no limitation on how many sources we could claim could be present in each cube, whereas here the methodology forbids claiming more than one source per cube. This means that if all 50 sources were detected, a maximum of 50 spurious detections could be claimed, which places an artificial limit on the minimum reliability possible : this cannot be less than half of the completeness, assuming the user always claimed the maximum possible number of spurious detections. This is shown by the solid line in figure \ref{fig:c-r}. A more realistic estimate of the lower limit on the reliability is shown by the dashed line, which assumes the user never claimed more than 60 sources were present in total (as per the actual statistics). For example if all 50 sources were detected, the number of spurious detections could not exceed 10. This is likely the result of the foreknowledge of approximately how many sources were present, and this constraint is at least partly responsible for the apparent correlation in figure \ref{fig:c-r}. However, there is no restriction on the maximum reliability, which could in principle be 100\% at any completeness level. The reason this is not so is a combination of both making fewer identification attempts when sources are harder to detect, (that is, at lower completeness levels) and in fewer of those attempts being successful, i.e. false positives.

Other biases are less important. Even for the sources injected with the highest line width, the fraction of each data cube occupied by a source was only 1\%, and for these cases the C/R values were close to 100/100. For the smallest sources the fraction was much less. We can therefore neglect the fraction of detections due to random chance. Importantly, it is evident from the table that the transition from low to high C/R values occurs over a very narrow range of line widths and S/N ratios. While the specific trend in figure \ref{fig:c-r} is at least partly a numerical artifact, more general conclusions regarding visual extraction capabilities should be secure.

\subsection{Variation in completeness and reliability}
\label{sec:comprelvar}
\begin{figure*}[t]
\begin{center}
\includegraphics[width=180mm]{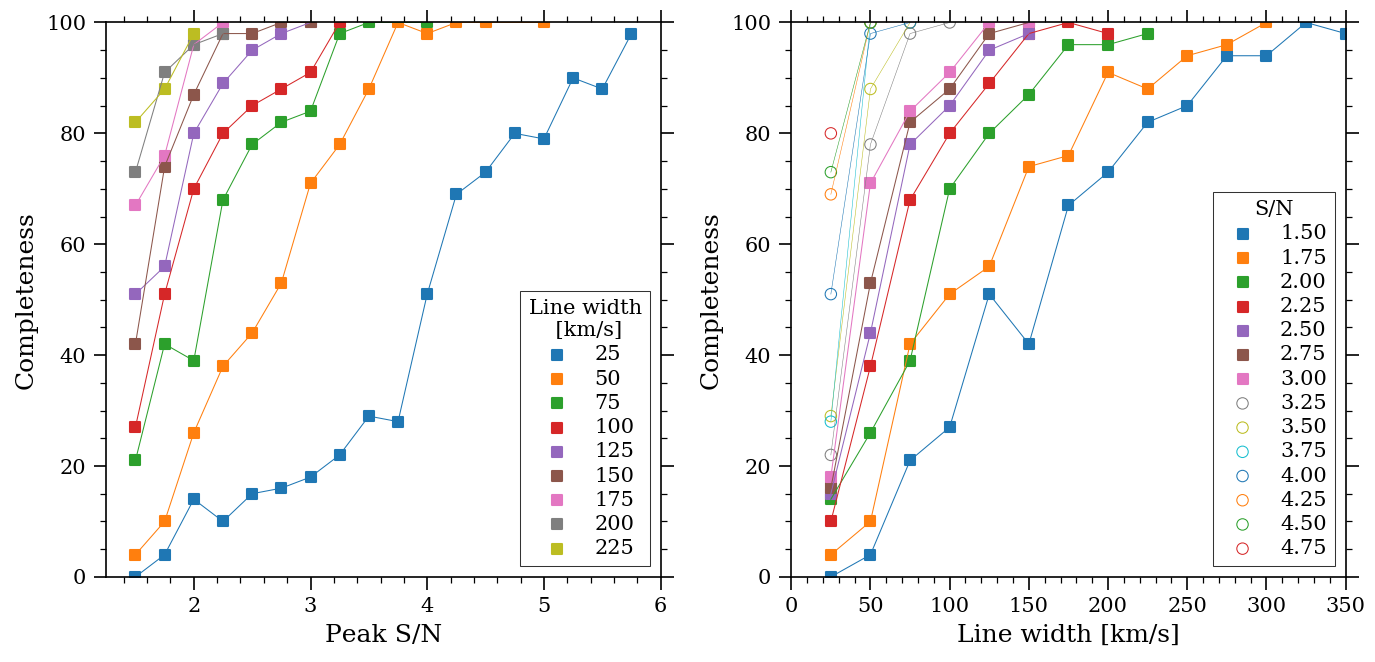}
\caption[c-sn-vw]{Completeness as a function of peak S/N (left) and line width (right), with one line per line width or peak S/N level respectively. The range of S/N and line widths has been truncated to avoid confusion, removing lines with only one or two data points; higher values simply show uniformly higher completeness levels.}
\label{fig:c-sn-vw}
\end{center}
\end{figure*}

Figure \ref{fig:c-sn-vw} shows the variation in completeness as a function of line width and S/N. Unsurprisingly, the higher either of these parameters, the greater the C/R values : sources which are either larger (at a given flux density) or brighter are easier to spot\footnote{The curves are not perfectly monotonic, with the occasional dip at increasing line width or S/N values. However the overall behaviour is clearly an increase in completeness with increasing S/N or line width, so the dips likely only represent the unavoidable imperfections in the visual extraction process.}. Sources above about 275\,\kms{} can be found with high confidence at even very low S/N levels because they are more noticeable as they span a larger region and because of \textit{noise boosting} (e.g. \citealt{multifind, boris}). That is, since the random variation in the background noise will mean some voxels are brighter than others, once even a weak signal is added to these, they may exceed the detectability threshold. The larger the source, the higher the chance that this will occur since there will be more bright noise pixels present anyway. Linear alignments of such features, however, seldom occur by chance, and coherent structures are readily visible to the eye against the random background noise.

Provided the line width is greater than about 50\,\kms{} (10 channels), features can be confidently identified as long as their S/N level exceeds 4.0 (importantly, these are the criteria we generally give in AGES papers as a sensitivity limit). Given that the C/R rapidly reaches 100/100 beyond this, this value is unlikely to be strongly affected by the numerical bias discussed above. Below this line width, a correspondingly high confidence is not reached until the S/N level reaches at least 6.0. Features this small resemble scarcely more than bright individual pixels, and even familiarity with the appearance of the features present is not enough to make them clearly visible to the eye. Detecting these smallest features might be improved by altering the data display range, but any gains would likely be marginal.

\begin{figure*}[t]
\begin{center}
\includegraphics[width=180mm]{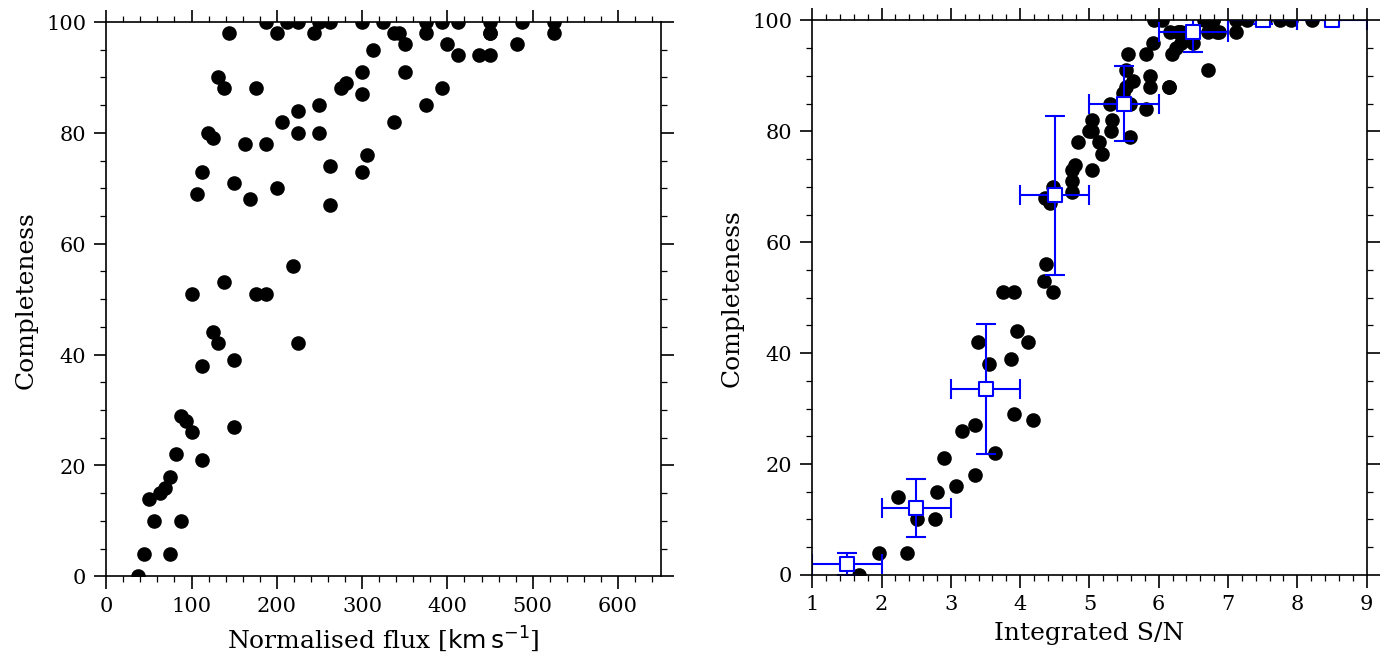}
\caption[c-fsn]{Left panel : completeness as a function of equivalent flux or mass. Right panel : completeness as a function of the integrated S/N parameter given in S07 and here in equation \ref{eqt:aasn}. The blue boxes show the median completeness level (with 1$\sigma$ vertical error bars) in bins of width 1.0 (indicated by the horizontal error bars) - see section \ref{sec:gocompare}.}
\label{fig:c-fsn}
\end{center}
\end{figure*}

\begin{figure*}[t]
\begin{center}
\includegraphics[width=180mm]{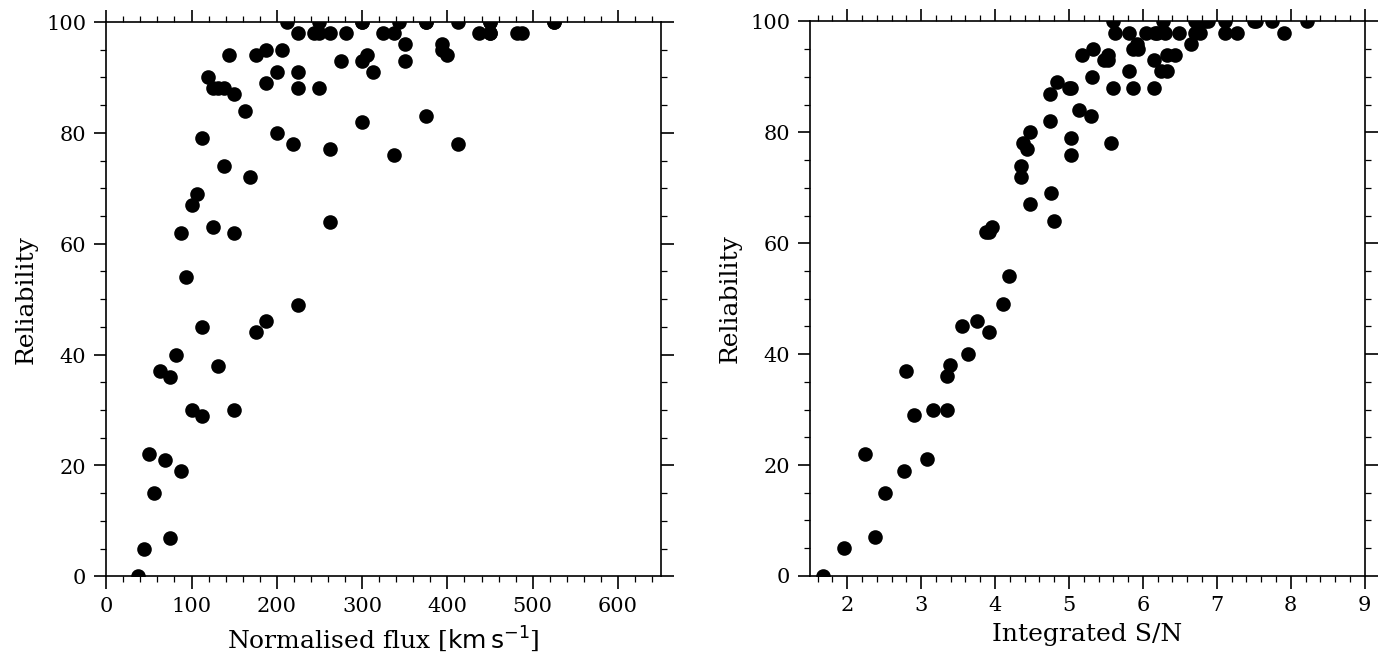}
\caption[r-fsn]{Left panel : reliability as a function of equivalent flux or mass in dimensionless units. Right panel : reliability as a function of the integrated S/N parameter given in S07.}
\label{fig:r-fsn}
\end{center}
\end{figure*}

The left panel of figure \ref{fig:c-fsn} shows the variation in completeness as a function of total flux (i.e. line width multiplied by S/N) for all sources, which at any given distance is directly equivalent to mass. There is a broad but highly scattered trend : at values of flux of 150 (here given in units of \kms{}, since the intensity values are normalised by the \textit{rms} but line widths must use physical units as per the equation below), completeness ranges from 20-100\%. This makes it all but useless as a survey parameter, though I will return to this in section \ref{sec:masscomp}. In contrast, the right panel in the same figure shows completeness as a function of the \textit{integrated} S/N parameter (hereafter S/N$_{\mathrm{int}}$); the corresponding plots for reliability are given in figure \ref{fig:r-fsn}. This is defined in S07 as follows :

\begin{equation}
\label{eqt:aasn}
S/N_{\mathrm{int}} =\frac{1000F_{c}}{W_{50}}\frac{w^{1/2}_{smo}}{rms}
\end{equation}
Where for W50 $<$ 400 km/s, $w_{smo} = W50 / (2 \times v_{res})$, where $v_{res}$ is the velocity resolution in km/s, and for W50 $>$ 400 km/s (though these values are not explored here), $w_{smo} = 400\,km\,s^{-1} / (2 \times v_{res})$. $F_{c}$ is the total flux in Jy\,\kms{}; $rms$ is the rms across the spectrum in mJy. In these experiments, since the flux is normalised by the \textit{rms}, the calculation of S/N$_{\mathrm{int}}$ simply omits the factor of 1000.

This parameter is routinely used throughout both ALFALFA and AGES papers. Here this plot shows for the first time that sources found by visual inspection follow a clear trend in this analytic formula originally derived in conjunction with S07's automatic source extraction algorithm developed for ALFALFA. The S07 prescription is therefore an excellent way to characterise what the eye is sensitive to, or equivalently, visual source extraction performance can be robustly and objectively quantified. Unlike S07, I do not differentiate between source properties (i.e. line width) in figure \ref{fig:r-fsn} because they appear to make little or no difference. Crucially, this parameter was never calculated ahead of time, so there was no possible foreknowledge of the source characteristics in this respect.

There is no particular reason to think that any of the numerical biases discussed in section \ref{sec:numberbias} would influence this result. One possible difficulty might be the choice of using uniform spectral profiles. \cite{profileshapes} show that only about a quarter of unresolved \HI{} detections have such a profile shape, while the majority of the others have either Gaussian or double-horns. This would certainly affect the accuracy of the results when trying to measure the source profile properties, but this is not the goal here. Any source can be approximated as a series of top hat profiles. Moreover, at any given total flux level, changing the profile from a top hat to \textit{any other shape of the same line width} must, by definition, \textit{increase} the integrated S/N over some parts of the profile. Therefore this is very unlikely to strongly affect the C/R results presented here.

S07's figure 7 shows their own estimates of completeness and reliability as a function of the integrated S/N. Interestingly, while their reliability shows very similar behaviour to figure \ref{fig:r-fsn} in the current work, completeness shows a much shallower rise, only approaching 100\% at much higher integrated S/N levels (10-12). We have used an integrated S/N level of 6.5 as a condition of reliability in previous works (the same as given in S07), however from the present study it appears that this is also an excellent value for completeness, in marked contrast to the S07 result. Their algorithm uses matched template filtering, with the injected sources having a more realistic distribution of spectral profile shapes, but they explicitly note that the alternative choice of using a top hat profile does not affect their results. 

One explanation for the difference in completeness levels of the S07 measurements compared to the results presented here may be the difficulties of measuring wider sources (high velocity widths pose problems for all automatic algorithms, which I will discuss further in section \ref{sec:gocompare}; see also section \ref{sec:aaowncomp} for detailed comparisons with ALFALFA). As S07 note, their completeness at the same S/N$_{\mathrm{int}}$ levels varies appreciably with velocity width. At S/N$_{\mathrm{int}}$ = 8.0, their completeness is approximately 65\% for sources of W\,$>$\,150\,\kms{} but over 90\% for sources with W\,$<$\,150\,\kms{}. In contrast the completeness from the visual extraction experiment here is close to 100\% at this S/N$_{\mathrm{int}}$, with a variation of less than one percent over all the sources measured.

\subsection{Mass completeness}
\label{sec:masscomp}
The mass of an \HI{} detection for a top hat profile is given by the standard equation :
\begin{equation}
\label{eqt:himass}
M_{\HI{}} = 2.36\times10^{5}\,d^{2}\,S/N\times\,W\,\times\,rms
\end{equation}
Where the mass is in solar units, $d$ is the distance to the source in megaparsecs, $W$ is the line width in \kms{} and $rms$ is the noise level of the data in Jy. The product $S/N\times\,W\,\times\,rms$ is the same as the total flux of the source. From this equation it is evident that at any given mass, the higher the value of $W$, the lower the S/N, and conversely, the higher the S/N, the lower the line width. Since both of these influence the detectability, as shown in figure \ref{fig:c-sn-vw}, defining a mass completeness limit is difficult. In principle, a source of any mass can be rendered undetectable by sufficiently increasing its line width, since this would lower its (peak) S/N. However it would also make the source larger and therefore -- naively -- one might expect it to potentially have a greater chance of being detected. These effects are in direct conflict with each other.

The result of figure \ref{fig:c-fsn} that visual detectability is strongly correlated with \snint{} allows us to resolve this conflict. We can use equation \ref{eqt:aasn} to examine how the detectability should vary as a function of mass and line width. At a fixed distance, by equation \ref{eqt:himass} flux is directly equivalent to mass. We can consider \textit{rms} to be a constant : here we only need consider idealised situations, but this is a good approximation to reality as most surveys have generally homogenous \textit{rms} levels. Velocity resolution is also well-approximated as constant. This means we can simplify equation \ref{eqt:aasn} :

\begin{equation}
\label{eqt:simpleaassn}
S/N_{int} \propto M / \sqrt{W} 
\end{equation}

Or equivalently :

\begin{equation}
\label{eqt:masslevel}
M \propto S/N_{int}\,\times\,\sqrt{W} 
\end{equation}

What this shows is that indeed mass completeness cannot be specified as a single number. At a constant S/N$_{\mathrm{int}}$ level (i.e. completeness or detectability), as the line width increases, so does the mass. This means mass completeness is width-dependent. Plots of the total flux as a function of line width in (for example) \cite{agesii}, \cite{agesiii} and \cite{boris} all show a clear tendency for verified sources to lie above the line of a constant S/N$_{\mathrm{int}}$ of 6.5, which compares well with the present finding that this is a good measure of survey completeness. While a small percentage of individual sources might be missed even at S/N$_{\mathrm{int}}$\,$>$\,6.5, this should not affect any claims about the statistical properties of a sample.

\begin{figure}[t]
\begin{center}
\includegraphics[width=90mm]{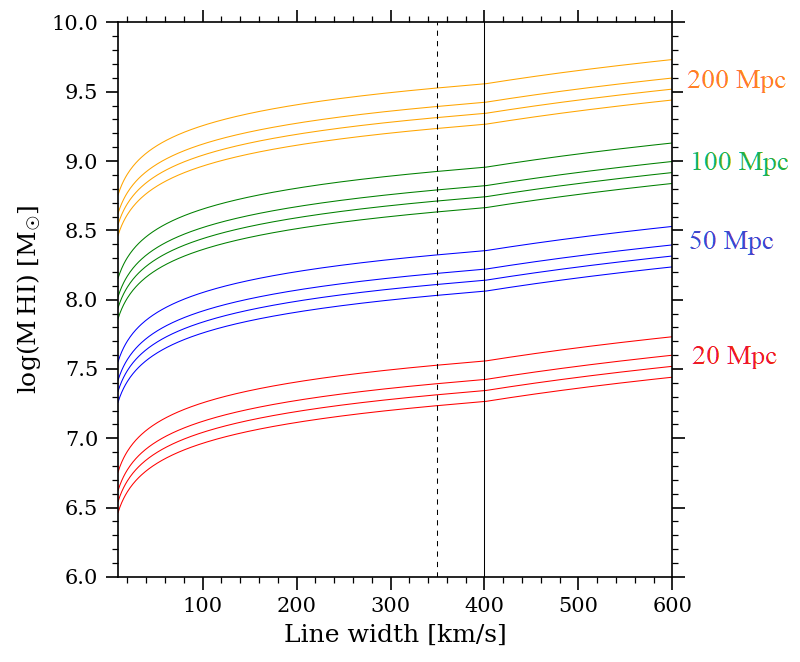}
\caption[masscont]{Mass-completeness contours for AGES at different distances. At each distance, the contours represent 25, 50, 75 and 100\% completeness levels, in ascending order. The dashed vertical line shows the 350\,\kms{} limit of the experimental results presented here, while the line at 400\,\kms{} indicates the break in the velocity smoothing function as described in S07.}
\label{fig:masscont}
\end{center}
\end{figure}

Another use of this is to parametrise how mass-completeness varies for a given survey at different line widths and distances. An example of this for AGES is shown in figure \ref{fig:masscont} (for the parameters of this and other surveys, see table \ref{tab:surveytable} and section \ref{sec:gocompare}). Here the completeness levels were estimated using the data for figure \ref{fig:c-fsn} by finding the median S/N$_{\mathrm{int}}$ at the specified completeness levels in bins with widths of 20\% in completeness. This emphasises that completeness is highly sensitive, varying at any line width from 25 to 100\% by only 0.3 dex in mass.

\section{Control tests}
\label{sec:biasproof}
As mentioned the experiment described above has several major differences from real data cube inspection, which may make this an unrealistic `best case' scenario for visual source extraction. Although the C/R values do reach 0/0, thus probing for sources so faint they are not detected even under these rather idealised conditions, it is entirely possible that the slope of the completeness curve is not realistic. For example, a more accurate curve might be steeper, truncated to zero at higher S/N$_{\mathrm{int}}$ values than seen here, or it might be shallower and not reach 100\% until a higher S/N$_{\mathrm{int}}$ value than shown in the curve of figure \ref{fig:c-fsn}. 

\begin{figure}[t]
\begin{center}
\includegraphics[width=90mm]{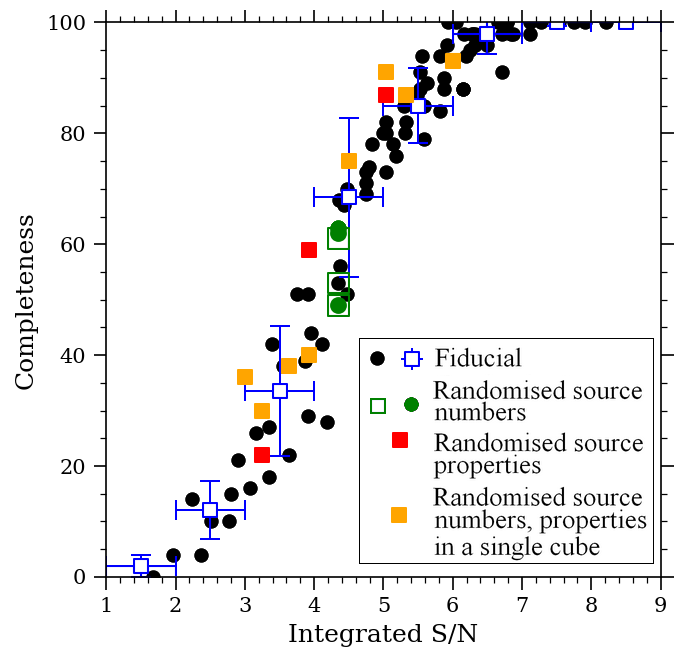}
\caption[biases]{Black points and blue squares are the same as in figure \ref{fig:c-fsn}, with other points from the experiments described in section \ref{sec:biasproof}. Specifically, open green squares describe the control test of section \ref{sec:foreprob}, which is run as in the fiducial case but at a precise S/N$_{\mathrm{int}}$ value of 1.945. The filled green circles show the corresponding tests at the same S/N$_{\mathrm{int}}$ value but with random numbers of sources injected, so that the observer had no knowledge of the probability a source was present. The filled red squares are as in section \ref{sec:seqtrain}, in which the observer had no knowledge of source line width or peak S/N and cubelets were inspected in random order, with no information given as to whether identifications were correct until the whole set of cubes was inspected. The filled orange squares are from the test described in section \ref{sec:quantise}, where numbers, line width, and peak S/N were all randomised at different S/N$_{\mathrm{int}}$ levels and sources were injected into a single large data cube rather than the standard set of 100 cubelets.}
\label{fig:biasproof}
\end{center}
\end{figure}

This experiment (hereafter referred to as the fiducial case) was optimised for speed to ensure a large sample size. That the results indicate a clear correlation between S/N$_{\mathrm{int}}$ and completeness, however, allows for a series of smaller but more carefully-controlled tests, in which both the statistical effects (i.e. the restriction to have at most only one source per cube) and the various possible sources of unconscious bias can be removed. The tests described below attempt to account for these concerns both individually and in aggregate. Results are shown collectively in figure \ref{fig:biasproof}.

\subsection{Foreknowledge of source probability} 
\label{sec:foreprob}
Although the user is never aware of the actual number of sources injected beforehand (indeed this is deliberately slightly randomised), they do know the probability that any cube contains a source is 50\%. This may make them more inclined to guess a source is present when they might otherwise suspect that this is not the case, and thus potentially increase the completeness level. 

I control for this as follows. First, I bin the fiducial results by completeness to find the median S/N$_{\mathrm{int}}$ which should give approximately 50\% completeness (3.93), which I randomised slightly to minimise foreknowledge still further (with the actual S/N$_{\mathrm{int}}$ value used for these tests being 4.35). I then perform three control tests using this precise S/N$_{\mathrm{int}}$ as this had not been previously examined. I arbitrarily chose a line width of 100 km/s for this test, with a peak S/N of 1.945. The control tests were run as in the fiducial case with full knowledge of the probability a source was injected. The resulting numbers of sources injected were 54, 49, and 51, with the corresponding completeness levels being 61, 53, and 49\% (open green squares in figure \ref{fig:biasproof}).

Next I ran three more inspections using the same source parameters but with the probability of injection being an entirely random number (0-100\%). In these examinations, the actual numbers of sources injected were 58, 67, and 84, with corresponding completeness levels of 62, 49, and 63\%. As shown by the filled green circles in figure \ref{fig:biasproof}, these values are entirely consistent with the scatter of the fiducial run at this S/N$_{\mathrm{int}}$ value. Foreknowledge of the probability of injection appears to be a very weak prior. The extreme cases of knowing a source definitely is or isn't present might conceivably be different. However the intermediate case of knowing the chance is 50\% amounts to little more than saying, "there might be a source present, or there might not" -- which is the assumption one has to make in \textit{any} real search; it gives the observer very little information on which to make a judgement.

There is little point in testing the case that the viewer knows a source is not present. The other extreme, that they know a source was injected, is more interesting. From table \ref{tab:bigtable} I selected inspections from the fiducial experiment where completeness was approximately 50\%, and within those I re-examined 16 cases where a source was present but I had not initially found. Re-inspecting these with the certain knowledge that a source was indeed present somewhere resulted in 6 additional detections but also 10 false positives. It appears that even certain foreknowledge does not play a strong role in source finding, at least in isolation from the other possible biases.

\subsection{Sequentiality and training} 
\label{sec:seqtrain}
In a real source extraction procedure, viewers may begin with the brightest, most obvious detections and then proceed to ever-fainter candidates. The bright sources act as references to the general characteristics of likely real sources, which viewers can use to visually extrapolate to anticipate the appearance of fainter objects. In addition, they will (or should !) have received extensive training with pre-examined data sets so they have a general familiarity with the appearance of real and spurious features in similar data. The fiducial experiment reproduces all of this, but might exaggerate these effects to a degree that influences the results. In a real search, viewers only have a much more limited control over what sort of sources they will look for next.

I account for this by injecting sources with randomised properties. As in the source probability test above, I bin the data by completeness but now select three different corresponding S/N$_{\mathrm{int}}$ values. Specifically, values of S/N$_{\mathrm{int}}$ of 3.25, 3.93 and 5.03 should have completeness values of 25, 50 and 75\% respectively -- assuming S/N$_{\mathrm{int}}$ is indeed the dominating factor rather than the foreknowledge of source properties and appearances.

This test modifies the fiducial experiment as follows. For each S/N$_{\mathrm{int}}$ value, 100 cubelets are created with a probability of source injection of 50\%. Individual source line widths are now randomised to the range 25 to 200 km/s. Unlike the fiducial case, the width is truly random within this range and not quantised into 25\,km/s intervals. For each width, peak S/N is calculated to give the required S/N$_{\mathrm{int}}$ value. Whereas in the fiducial case sources are in directories named according to their line width and peak S/N values, here the file and directory names use a random alphanumeric string. They are inspected in random order, and when the user makes their determination as to the presence of a source, they are given no information on their success until the end of the whole experiment (i.e. after inspecting all 300 cubelets). In short, for each source the user inspects, they have no prior knowledge of its detectability (varying from very unlikely to highly probable), whether a source is present at all, or its properties if it was injected (that is, its visual appearance). This is closely analogous to real searches where the user has no foreknowledge of source properties save their previous training.

The results are as follows :
\begin{itemize}
\item In the S/N$_{\mathrm{int}}$ = 3.25 bin, 46 sources were injected and the completeness was 22\%
\item In the S/N$_{\mathrm{int}}$ = 3.93 bin, 39 sources were injected and the completeness was 59\%
\item In the S/N$_{\mathrm{int}}$ = 5.03 bin, 54 sources were injected and the completeness was 87\%
\end{itemize}
As shown by the filled red squares in figure \ref{fig:biasproof}, these results are again entirely consistent with the general scatter in the original experiment. Foreknowledge of the source properties and inspecting sources with sequentially-modified parameters does not appear to much affect the detectability rates.

\subsection{Quantising the data into cubelets}
\label{sec:quantise}
In the fiducial case, one knows that there either is or isn't, at most, one source present per cube. In a real data set, sources may be present (in principle) in unlimited numbers everywhere. That is, when a source has been found, the observer must still keep searching the same nearby area and can't rule out that many more sources may be present just because a single object has already been found there.

I address this issue by injecting large numbers of sources into a single data cube. As noted in section \ref{sec:setup}, the disadvantage of this is that is requires sources to be masked to avoid confusion. This would make an experiment on the scale of the fiducial case unviable, but for smaller tests is quite achievable. As in the above test, I inject sources of S/N$_{\mathrm{int}}$ values 3.25, 3.93, and 5.03, for expected completeness rates of 25, 50, and 75\%. Again, source properties are randomised with velocity widths ranging from 25 to 200 km/s and peak S/N levels calculated to give the required S/N$_{\mathrm{int}}$ value. Unlike the previous test, I now set the source numbers in each bin to a random number between 0 and 100 (the upper limit being necessary to avoid too high a source density and also to ensure the test can be carried out in a practical time). This number is not known to me until the search is completed. Furthermore, I use a different data set than in the fiducial case. I extract a cube from the same Mock spectrometer data of NGC 7448, but now expand the velocity range to -7,000 km/s to -1,000 km/s. This means I search an area I have never before inspected at all.

This is about as close to reproducing a real search as can be done. Since masking the sources is required, unlike the fiducial case I now inspect the data in 3D volumetric display mode before moving to 2D image slices. I run this test three times, each time with slight modifications. The results are summarised below and shown as the filled orange squares in figure \ref{fig:biasproof}, though omitting the meaningless case where (since sources numbers were randomised) only a single source was injected:
\begin{itemize}
\item The first test proceeded exactly as described above. In ascending order of injected S/N$_{\mathrm{int}}$, 89, 25 and 11 sources were injected, with corresponding completeness values of 30, 40 and 91\%. The number of false positives was 29, for an overall reliability of 62\%.
\item The second test randomised the S/N$_{\mathrm{int}}$ values slightly to 3.63, 4.35, and 5.23. The number of sources injected in each bin was 8, 1, and 97, with completeness rates of 38, 100 (but this is statistically meaningless as only such one source was injected) and 85\%. The number of false positives was 30, giving an overall reliability of 75\%.
\item For the third instance I increased the maximum number of sources allowed at each S/N$_{\mathrm{int}}$ value to 300 for better statistics. I also deliberately picked S/N$_{\mathrm{int}}$ values to ensure a fuller sampling of the completeness curve: 3.0 (22 sources), 4.5 (72 sources) and 6.0 (159 sources). The completeness rates were 36, 75, and 93\%. The number of false positives was 45, for an overall reliability of 82\%.
\end{itemize}

Despite the large differences in source densities, these numbers are all consistent with the original completeness curve. It appears that none of the potential biases considered have unduly influenced the original result. Crucially in this case, to try and ensure a high completeness rate for the faintest sources, I had to assume that essentially all the faintest structures visible were viable candidates. This is quite different to the fiducial case where one knows to ignore very weak structures when searching for the brightest objects. Yet this did not result in a dramatic difference in reliability or inordinate numbers of false positives.

In the end, it appears that the eye sees what the eye sees : one has to judge each potential source on its own merits, and foreknowledge of what may or may not be present has remarkably little effect on this.

\subsection{Seeing is not necessarily believing} 
\label{sec:noticing}
Throughout all of the above experiments, the `ground truth' -- the full, true source population -- is known with certainty, and false positives can be verified as such with equal certainty. This is a luxury not available for observational data. Conceivably, observers under real conditions might be forced to reject more of the weaker sources than are recovered in the experiments presented here. They may well notice them at some level, but choose not to investigate or record them on the grounds of needing to reject false positives that would otherwise degrade the accuracy of their findings (or require additional telescope resources to verify). This `recording bias' might mean the completeness rates found here are artificially higher than if the same experiments were run on genuine science data.

\subsubsection{How AGES verifies detections}
The magnitude of this effect depends on two factors : how astute observers are at initially recording candidate sources, and their methods of verifying those candidates. This will necessarily be survey and observer-dependent. In AGES, our strategy in our initial searches has been explicitly to make source lists which are as complete as we can, in essence accepting any plausible-looking source. We do this because one of the primary aims of the survey is to achieve high sensitivity, as required for detecting dwarf galaxies and other objects in more distant environments, but also because it is far easier to vet a long source list than having to redo the source extraction completely, which would be necessary if we suspected we were missing a substantial population of faint sources. 

Numerous methods of validating the initial catalogues are used. The AGES standard is to either have multiple observers inspect the same data set or run an independent search using automatic algorithms (or both). Independent detection provides strong evidence that something is present which is worth investigating further, though a source only found by one algorithm or observer may still be real. We can also inspect the data gridded in individual polarisations, which can reveal spurious sources as these are sometimes significantly stronger in one polarisation than the other. This too, though, is not a guarantee, as while polarised sources can be rejected, weak sources may still be spurious even if unpolarised. 

For true verification we have two options. One is use to data from other wavelengths, primarily optical. A weak source with an optical counterpart close to the \HI{} spatial coordinates may be considered more probable as real emission, especially if it has a matching optical redshift. This is a powerful way of validating sources, with \cite{nodarks} finding that only 1.5\% of sources lack optical counterparts (conversely, \citealt{harvesting} find that candidate detections without optical counterparts are almost entirely spurious). This is not in any way to diminish their astrophysical interest, but in pure numerical terms optically dark \HI{} sources (at least for unresolved, extragalactic emission) are clearly not significant. Our other option is to conduct follow-up 21\,cm observations, for which we have generally used the L-wide or ALFA receivers on Arecibo. Since follow-up observations were included in the allocated AGES observing time, we were able to make liberal use of this, re-observing sources whenever there were any grounds at all to be suspicious of a source -- primarily this was based on S/N, but could also include relatively bright detections if the optical data did not reveal an obvious counterpart. Our published catalogues are therefore almost entirely limited to verified real signals (follow-up for ALFALFA is discussed in detail in \citealt{harvesting}).

The inspection techniques and verification methods give some hope that the recording bias against faint sources is of limited impact. Not only do we explicitly train observers to search for faint sources, but the tools offered in FRELLED make vetting the sources straightforward and rapid. In particular we routinely query the Sloan Digital Sky Survey (SDSS) to search for optical counterparts, since one of our goals is to search for optically dim and dark \HI{}, as well as inspecting the quality of the spectra (typically in both polarisations). Finally, as mentioned in the introduction, our follow-up observations have had reliability levels as low as 28\%, indicating that we are indeed probing deep into the noise, and the S/N$_{\mathrm{int}}$ distribution of our sources does probe below the putative 6.5 reliability threshold, as will be discussed in the next section.

\subsubsection{Control tests for the recording bias}
\label{sec:noticetests}
I performed further tests to validate this finding. In the previous experiments I selected a master cube known to be devoid of real signals. In this final control test, I use cubes in which real signals are known to exist. For this I selected the VC1 (\citealt{agesvc1}) and Abell 1367 cubes (\citealt{boris}), in both cases selecting the velocity range 10,000 -- 15,000\,\kms{} as the source density there is relatively high. Using our published catalogues, I first masked the known real sources, then injected artificial sources randomly throughout the remaining volume in the manner of the experiment described in section \ref{sec:quantise}. The number of sources was randomised with a maximum of 100 and the \snint{} set to 3.93, known to give expected completeness rates of approximately 50\%. The primary goal here was not to measure completeness rates as in previous tests, but instead to consider the false positives found during this new search. If this new extraction recovers the artificial sources at the typical completeness rates, and if there had been a strong recording bias against faint sources during the initial searches of these cubes, then the false positives (that is, sources which were not injected as part of the experiment) ought to contain significant numbers of plausible real, astrophysical candidates. 

In the VC1 volume considered there were 103 known real sources present, while in Abell 1367 there were 123. These high numbers are important to control for the effects of large scale structure : to give the best chance of finding any additional faint sources, we need to search areas where galaxy density is known to be high (though both volumes include velocity ranges where the galaxy density is much lower than others). I searched these cubes in the same manner as in the previous tests, relying exclusively on the \HI{} data for the initial catalogue, but this time vetting the resulting list using a combination of the injected source catalogue and optical SDSS data. I found that while completeness rates for the artificial sources were, once again, consistent with all the previous experiments, the number of plausible new real, astrophysical sources (that is, having a plausible optical counterpart, i.e. a visible galaxy near the \HI{}) either with an optical redshift in agreement with the \HI{} or without any optical redshift data) was extremely small : 6 for VC1 and 7 for Abell 1367, i.e. 5\% of the previously-known totals. I stress that these are also likely upper limits, with the verification here limited to visual inspection of the optical data.  

In short, the role of the recording bias appears to be minor. Whilst I here recovered faint sources in agreement with the completeness rate determined in the fiducial experiment, this same search revealed very few additional candidate real galaxies. Thus the completeness curve of figures \ref{fig:c-fsn} and \ref{fig:biasproof} is likely to be a reasonably accurate reflection of that obtained during the actual AGES and WAVES surveys. The results for VC1 are particularly encouraging, given that this region was previously only searched using \textit{kvis}.

\section{Comparisons to algorithmic source extraction}
\label{sec:gocompare}
The fiducial experiment described in section \ref{sec:results}, together with the supplementary tests of section \ref{sec:biasproof}, establishes that visual extraction is hardly an entirely subjective procedure which suffers from the whims and unpredictable moods of the observer. Instead its efficacy can be readily and precisely quantified. While this information is of merit by itself, it is also beneficial to compare these results with those from automatic source extraction algorithms. I do this in two ways. First, I run two algorithmic source extractors on the data sets used for the visual source-finding experiments. This has the advantage of running the tests under (essentially) identical controlled conditions, thus giving results for each which are directly comparable. Secondly, I also compare the results of published catalogues from surveys which use different source extraction methodologies. This is less rigorous, but has the value of using the different procedures under truly real-world conditions.

Algorithmic source-finders can be used in different ways. Users can, for example, simply set the search parameters and choose to accept the results without any further involvement. These fully automatic catalogues are suitable for scientific analysis provided prior testing establishes they have sufficient C/R values, even if not necessarily at the ideal levels of 100/100. Alternatively, and more frequently, algorithmic extractors may be used in combination with visual extraction. Different kinds of such hybrid mode are possible. For example, one or more human observers may search the cube and complement their results with a fully-independent algorithmic search, or they might pass a masked cube to the algorithm so that it only searches regions where visual inspection did not find any candidate sources. Perhaps the most common hybrid approach is to rely on the algorithm to generate the initial candidate list and restrict human involvement to a vetting role. This greatly reduces the burden of work for the observer since they do not have to search the entire data cube, but also reduces the scope of visual inspection to find sources the algorithm missed.

Regardless of how automatic extractors are used in actual science analysis, they are usually tested by running them as independent tools. That is, whether their candidate lists are compared with the detection rates of known astrophysical sources or artificially-injected signals, completeness and reliability are typically quoted for how the extractor performs on its own without human vetting of their candidate lists (e.g. S07, \citealt{popfind}, \citealt{SOFIA2})\footnote{Vetting of the candidates can only reduce the completeness of the \textit{initial} list, so the completeness levels obtained herein will be upper limits. The effect on reliability may, however, be more complex, and use of visual verification may encourage users to deliberately use search parameters which give initial results of low reliability but high completeness.}. Given the various methods available for using algorithms in combination with visual extraction, I here follow the standard procedure and restrict the following performance estimates to the purely automated cases. I stress, however, that these hybrid modes can be valuable, and indeed we generally use automated extractors for AGES and WAVES as a way of complementing our visual searches.

\subsection{Direct comparisons : SoFiA, GLADoS, and visual source-finding}
For the direct comparisons I select the source extractors SoFiA and GLADoS. SoFiA is an extremely popular and versatile tool which is widely used in \HI{} studies, while GLADoS was designed specifically for AGES data and has not (to my knowledge) been used elsewhere. As described in \cite{SOFIAreliable, SOFIA}, SoFiA works by using a ‘smooth and clip’ algorithm, applying spatial and spectral smoothing to the cube on multiple scales and searching for emission above specified thresholds at each smoothing level. Reliability is assessed by comparing the properties of each potential detection with the statistics of the negative values of the noise, i.e. rejecting candidates if an equivalent ‘detection’ can be found at negative flux levels. 

GLADoS is a simpler algorithm which searches for pixels above a specified peak S/N threshold in structures which exceed a given velocity width threshold. Its search proceeds along all spectra in the cube and all candidates within a given spatial radius and velocity width are merged, with sources rejected if they were detected in less than a given number of pixels. Optionally, GLADoS can also check the detections in individual polarisations (since the \HI{} signal is unpolarised and the noise between the two polarisations should be uncorrelated) to eliminate polarised sources, though this is not used here as data for the individual polarisations is unavailable.

The efficacy of GLADoS was quantified in \cite{agesvc2}, but rather crudely : completeness and reliability were estimated based on a catalogue of real sources in the AGES VC2 field as generated by all source extraction methods employed there (visual plus other algorithms). SoFiA's performance was previously estimated in a manner much more similar to the visual experiments performed here, by having it search cubes into which artificial signals were injected (\citealt{SOFIA2}). Neither was previously quantified in the manner of the visual experiments described in this work.

To compare the automatic and visual methods, we need to be able to measure completeness and reliability as a function of \snint{}. Extracting sources from a single cube, which was shown in section \ref{sec:quantise} to give the same results as in the fiducial case, is far easier for the automatic algorithms than running on many cubelets, especially given the need for experimentation to find the ideal parameters and testing the results on sources of precise \snint{} levels. I therefore adopt a similar procedure to that of section \ref{sec:quantise} but with modifications to account for the automatic extractors. Both algorithms use the same basic setup, with some minor modifications for each.

\subsubsection{Experimental setup}
The source cubes are created as follows. Over a range of \snint{} from 1.0 to 10.0 in steps of 0.1, for each \snint{} value exactly 100 sources are injected into the standard master noise cube, with randomised velocity widths from 25 to 200\,\kms{}. As in section \ref{sec:quantise}, peak S/N is calculated to give the specified \snint{} given the selected velocity width, and sources are injected into the raw cube before Hanning smoothing is applied. Sources are here distributed in a uniform grid pattern, as preliminary testing found that the algorithms could sometimes merge sources if they happened to be too close together. The central spectral channel of each source is randomised, so that sources are always found throughout the entire volume of the cube. Whereas a human would very quickly realise that sources were distributed in a grid structure, this is irrelevant for the algorithms used here.

The SoFiA results are generated entirely automatically, with a script injecting the sources, running the search, comparing the results with the injected source catalogue, and calculating completeness and reliability. The completeness and reliability dispersions are calculated in exactly the same way as for the fiducial case, by binning the results in \snint{} intervals equal to one (centred on 1.5, 2.5, etc.). The median and dispersion values are thus calculated from ten searches of a total of 1,000 sources in each bin.

As an interactive program, GLADoS is more difficult to modify to allow full automation, and is also much slower, making a search of the same fidelity as possible with SoFiA impractical. Instead, the injected \snint{} values used for GLADoS were set to be from 1.5 to 9.5 in steps of 1, i.e. the same bin centres used for SoFiA, in each case again with exactly 100 sources injected with the same properties as used for the SoFiA search. GLADoS was then run manually on each injected cube (with the same search parameters in each case).

\subsubsection{Search parameters}
\label{sec:searchpar}
As mentioned in section \ref{sec:noticing}, for both real AGES searches and the investigations performed here, completeness is favoured over reliability. However when comparing the results with the algorithms, we need to compare both completeness and reliability together -- 100\% completeness is of no use if reliability is 0\%. Prioritising one over the other is legitimate, but in any practical situation, we cannot entirely neglect either performance parameter.

The approach I adopted here was to attempt to calibrate the search parameters for each algorithm using sources at \snint{} = 3.93, the level at which the fiducial case gave a median completeness of 50\%. Once parameters were found which gave results in agreement with this value, I ran the search on the whole set of \snint{} values as described above, enabling the completeness curves to be plotted. I also investigated other parameters in an attempt to maximise reliability and/or boost completeness beyond the 50\% achieved in a visual search, as will be described below. 
	
GLADoS has only a very limited set of parameters which can be set, principally the peak S/N threshold and velocity width. In real searches we have generally used values of 4.0 and 20.0\,\kms{} respectively. Through trial and error, slightly better results were found here using values of 3.75 and 25.0\,\kms{}, which gave C/R values of 32/26. This is well below the human completeness level of 50\% even allowing for the strong scatter in the measurements, but this is unsurprising : GLADoS was designed for a hybrid approach to source finding rather than being a wholesale replacement for visual extraction, and has not been previously tested on large numbers of faint sources in this way.

SoFiA has a much larger set of possible input parameters. After testing these extensively, the following parameters were found to give C/R values of 59/31 (this completeness value is slightly higher than the visual result though well within the general scatter), with others left unset :
\begin{lstlisting}
scfind.kernelsXY = 0
scfind.kernelsZ = 0, 3, 5, 7, 9, 11, 13, 15, 17, 19, 21, 23, 25, 27, 29, 31, 33, 35, 37, 39, 41
scfind.threshold = 3.5
linker.radiusXY = 2
linker.radiusZ = 3
linker.minSizeXY = 3
linker.minSizeZ = 4
reliability.enable = false
\end{lstlisting}

Setting \textit{scfind.kernelsXY} to zero disables spatial smoothing, which is not needed here as all sources are unresolved. The parameter \textit{scfind.kernelsZ} sets spectral smoothing to use all possible channel widths up to the widest sources injected. The \textit{scfind.threshold} parameter is effectively the peak S/N threshold, while the linker parameters control the merging, here ensuring the sources span one beam and are at least 20\,\kms{} in spectral extent. The reliability module is disabled, as testing found this would always cause a substantial decrease in completeness. Similarly, the \textit{scfind.threshold} value was found through fine-tuning. While the final value of 3.5 gave C/R values of 59/31, this is highly sensitive to changes. A value of 3.25 increased completeness to 69\% but with a dramatic drop in reliability to just 13\%, with a further drop to 3.0 reducing completeness back to 59\% with reliability down to 5\% (likely due to merging of sources because of the larger number of false positives). Increasing it improved reliability but at the cost of completeness : a value of 3.75 gave C/R\,=\,39/53, while a value of 4.0 gave C/R\,=\,35/60.

\subsubsection{Results}
\begin{figure}[t]
\begin{center}
\includegraphics[width=90mm]{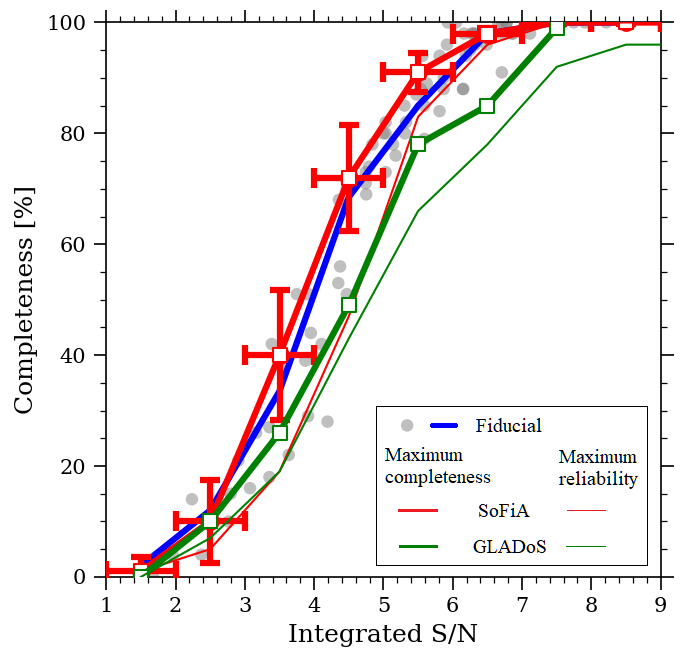}
\caption[r-fsn]{Comparison of the results of the fiducial visual source extraction experiment described in section \ref{sec:results} with automatic algorithms, showing completeness as a function of \snint{}. The grey points and blue line (here omitting the error bar for clarity - see figure \ref{fig:c-fsn}) shows the fiducial visual experiment, while SoFiA is shown with red lines and GLADoS by green lines. Thick lines show the results after optimising the algorithms to maximise completeness whereas the thin lines are attempts to maximise reliability.}
\label{fig:sfcomp}
\end{center}
\end{figure}

The completeness results are shown in figure \ref{fig:sfcomp}. Thick lines show the results for the tests described above, which attempted to maximise completeness. All three completeness curves are qualitatively similar. It is immediately apparent that SoFiA (red) does equally well in terms of completeness as visual extraction (blue line and grey points) at all \snint{} levels, even slightly exceeding it though within the general scatter of the visual results. GLADoS (green) does significantly worse, still just about within the scatter of the visual searches but clearly inferior to SoFiA.

Completeness is straightforward to estimate from the experiments described, but reliability is more ambiguous. The number of false positives found by any algorithmic search is essentially independent of the arbitrary number of real sources present (neglecting cases of overlap). Moreover, for the faint sources considered here, both GLADoS and SoFiA give highly scattered measurements of velocity width and flux that make their \snint{} estimates essentially meaningless (though both do better when brighter sources are used). A simpler and more robust way to compare the relative reliability of these procedures is to consider the number of false positives found in the same search volume. From the earlier tests, the highest number of false positives found visually in this region was 19, whereas SoFiA found 133 and GLADoS 176 spurious detections.

These results show that SoFiA can reach completeness levels comparable to visual extraction but at a significantly lower reliability, by a factor of seven. This is, however, considerably better than GLADoS, which does not reach the same completeness as visual extraction even while being a factor of nine times less reliable. It should be noted that the number of false positives (i.e. reliability) is strongly dependent on data quality, and these quantitative results will not translate directly to other surveys.

I also tried fine-tuning the input parameters for the algorithmic extractors to optimise for reliability rather than completeness, i.e. setting them to obtain approximately the same number of false positives in the same area as found by visual inspection. For SoFiA, the closest result was obtained by setting \textit{scfind.threshold} to 3.80 (with the other parameters as in section \ref{sec:searchpar}), giving 22 false positives. For GLADoS it proved impossible to achieve a comparable result, so instead the standard search parameters (peak S/N\,=\,4.0, velocity width 20\,\kms{}) were used, giving 93 false positives. These results are shown by the thin lines in figure \ref{fig:sfcomp}. The drop in completeness for SoFiA is substantial, reducing from 72\% at \snint{} \,=\,4.5 to 52\%. However this is no worse than GLADoS at the same level, and even in this case it still converges to $\approx$ 100\% completeness at \snint{}\,=\,6.5.

By any metric, SoFiA is clearly superior to GLADoS : it achieves better completeness and reliability in a fraction of the time, taking of order 30\,s for the standard master cube whereas GLADoS requires several minutes. Comparing SoFiA and visual extraction gives a more ambiguous result. While SoFiA does not reach the same balance of completeness and reliability as visual extraction, the major advantage of speed should not be ignored. For instance, we have not found even the poorer results of GLADoS to be an impediment to its use because it delivers a catalogue much faster than visual inspection, with that catalogue being not just a list of source positions but also optical data, moment maps, and PV diagrams. Likewise, while the completeness-optimised use of SoFiA is substantially worse in terms of reliability than visual inspection, in practical terms this may not actually matter for its use in a hybrid mode : the large size of the catalogue list will be offset by the speed and convenience of examining its results compared to viewing the raw data cube. 

It is interesting to note that both SoFiA and visual inspection achieve close to 100\% completeness at a \snint{} level greater than about 6.5, even at very similar overall reliability levels. While visual inspection reaches, compared to the reliability-optimised SoFiA, significantly better completeness, this is only true for the faintest detectable sources (i.e. those below \snint{}\,=\,6.5). This suggests the following strategy for optimising source extraction efficiency. For bright sources of \snint{}\,$\geq$\,6.5, SoFiA alone is sufficient. For fainter sources, the situation depends on the survey size. Visual extraction achieves a better C/R balance, but for larger surveys, it may still be more efficient to accept a lower reliability from SoFiA and only use visual inspection to verify SoFiA's candidates. Finally for smaller surveys, visual extraction alone is sufficient, and combining the results with SoFiA is unlikely to significantly increase completeness levels. Searching in the third test described in section \ref{sec:quantise}, SoFiA found 170 sources (compared to 212 for visual inspection), of which only 10 were missed visually. 

The major caveat to this conclusion is that the SoFiA results were deliberately fine-tuned, which is not possible in real data where ground truth is not known (though \citealt{MeerGPS} calibrate their input parameters based on a visual inspection). Furthermore the number of parameters available for SoFiA users is large, and whether users obtain completeness results as high as here will depend on their particular use cases. Conversely, visual extraction is undeniably more laborious and requires both extensive training and familiarity with the data sets to achieve these results.

\subsection{Survey comparisons}
\label{sec:comparethemeerkat}
For the second method of comparing source extraction methods I used catalogues produced by a selection of different surveys. Estimating completeness from survey catalogues is routinely used in \HI{} surveys, e.g. \cite{hideep}, \cite{HiComplete}, \cite{AA40}, primarily for considering measurements of the \HI{} mass function. For evaluating source extraction techniques, this is necessarily less rigorous than using the carefully controlled conditions of the same noise cube with known injected sources, but has the advantage of using truly real-world conditions. I selected a set of surveys for this comparison whose characteristics are summarised in table \ref{tab:surveytable}. These were chosen based on the availability of catalogued data and their having a significant number of detections, at least a few hundred. The catalogues were obtained from the following publications : ALFALFA data is from \cite{AA100}; FASHI (FAST All Sky \HI{} Survey) is from \cite{fasthi}; HIPASS (\HI{} Parkes All Sky Survey) is from a combination of \cite{hipassmain} and \cite{hipassnorth}; and AGES data uses all source catalogues published so far - \cite{ages}, \cite{agesii}, \cite{agesiii}, \cite{agesiv}, \cite{agesvc1}, \cite{agesvc2}, \cite{agesvii}, and \cite{boris}.

\begin{table}[t]
\tiny
\begin{center}
\caption[surveytable]{Summary of selected \HI{} survey parameters.}
\label{tab:surveytable}
\begin{tabular}{c c c c c c}
\hline
\multicolumn{1}{c}{Survey} &
\multicolumn{1}{c}{Spatial} &
\multicolumn{1}{c}{Spectral} &
\multicolumn{1}{c}{\textit{rms}} &
\multicolumn{1}{c}{Extraction} &
\multicolumn{1}{c}{No. sources} \\
& res. & res. & & method & \\
& [arcmin] & [km/s] & [mJy] & & \\
\hline
ALFALFA & 3.5 & 10 & 2.3 & Automatic & 31,502\\
AGES & 3.5 & 10 & 0.6 & Visual & 1,231\\
HIPASS & 15.5 & 18 & 12.2 & MULTIFIND & 5,317\\
FASHI & 2.9 & 6 & 1.5 & SoFiA & 41,741\\
\hline
\end{tabular}
\tablefoot{The \textit{rms} is quoted from the median level of all detections in each survey at its nominal resolution, before any additional smoothing was applied.}
\end{center}
\end{table}

In brief, the source extractors used by the other surveys are as follows. ALFALFA use their own algorithm as described in S07, which uses a matched-filtering technique using a variety of template spectra that accounts for varying width and spectral profile shapes. HIPASS used the \textsc{multifind} (\citealt{multifind}) and \textsc{tophat} (\citealt{hipassmain}) algorithms. \textsc{Multifind} detects pixels above a peak S/N threshold, assuming a constant median noise level and requiring sources to be present in adjacent voxels. It also searches the cube at multiple levels of spectral smoothing. FASHI use the SoFiA extractor. 

\begin{figure*}[t]
\begin{center}
\includegraphics[width=180mm]{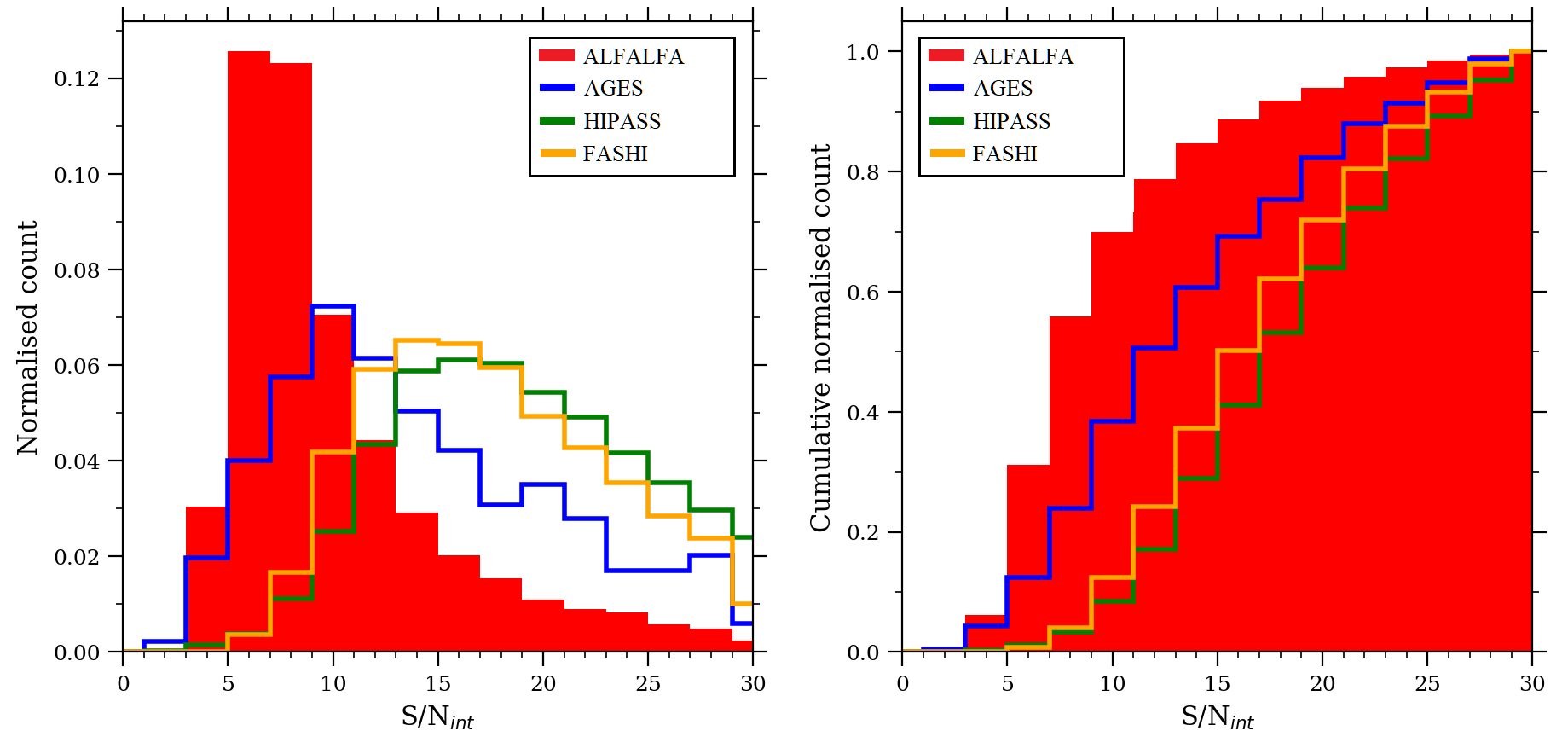}
\caption[r-fsn]{Comparison of the integrated S/N distribution for a selection of \HI{} surveys as indicated in the legend. The left panel shows the normalised count while the right panel shows the cumulative distribution. The horizontal range has been truncated to values $<$\,30 in order to show the shapes of the distribution more clearly. Note that these S/N$_{\mathrm{int}}$ values, while calculated using the original survey data, do not present a fair comparison for assessing source extraction methods : see figure \ref{fig:allcorrect} for a more accurate comparison.}
\label{fig:sndistro}
\end{center}
\end{figure*}

Given the very clear correlation between S/N$_{\mathrm{int}}$ and completeness in figure \ref{fig:c-fsn}, the integrated S/N is an appealing parameter to use for comparisons, especially since it accounts for different \textit{rms} and velocity resolution levels, and is almost independent of the shape of the spectral profile. In the following analysis, I used the S/N$_{\mathrm{int}}$ provided directly in the ALFALFA and FASHI catalogues, while for HIPASS I computed this using the data given in their published tables. The raw distribution of the S/N$_{\mathrm{int}}$ values for the different surveys is shown in figure \ref{fig:sndistro}. 

Immediately obvious from this comparison is that ALFALFA find a much higher fraction of their sources at lower \snint{} levels than the other surveys. Interpreting the differences between the completeness curves requires caution, however, as the surveys are of unavoidably different natures. Not only do their spatial and velocity resolutions vary (which will affect how sources with the same intrinsic properties appear in the different data sets), but so do their sky and spectral coverage areas. In particular, a survey of lower sensitivity may still find more faint sources than a deeper survey, if there are simply more such sources present in its search area for it to find -- even if its extractor has lower completeness as a function of S/N. Additionally, the \snint{} parameter means that the same sources will be detected at different levels in somewhat subtle ways. The effect of the \textit{rms} is straightforward : the better the sensitivity, the higher the \snint{} of any given source. However the \snint{}, as per equation \ref{eqt:aasn}, also includes the velocity resolution of the survey, which has a less obvious effect on the measured S/N of a source. 

In reality, to achieve the same \textit{rms} level at a better velocity resolution \textit{using the same telescope} would require a longer integration time, thus these parameters are not strictly independent even though they are treated separately in equation \ref{eqt:aasn}. However when comparing different surveys as here, the parameters are indeed fully independent because antenna size also plays a crucial role in sensitivity. Even on the same telescope, there are many other factors affecting \textit{rms} besides velocity resolution, e.g. changes in the antenna shape due to thermal variation or movement of the dish or receivers, signal leakage from electronics, RFI, elevation of the source, and weather (though usually only in extreme cases for \HI{}). By treating \textit{rms} and spectral resolution independently, equation \ref{eqt:aasn} calculates \snint{} using the actual observed parameters, and so intrinsically accounts for all of these possible factors.

This means that for any given source, if we allow (purely for the sake of exploring the consequences of the \snint{} calculation) the \textit{rms} to remain the same but improve the velocity resolution, the \snint{} will necessarily increase. The effect of this is different to changing the \textit{rms} by itself. Improving the \textit{rms} would increase the \snint{} of all sources, leading to an increased number of detections of sources of lower flux. The same would not happen for velocity resolution. As discussed in section \ref{ref:data}, there is a lower limit on the width of the \HI{} line itself, with none expected to be narrower than 10\,\kms{}. Improving the resolution beyond this, as for example in the case of FASHI, will therefore boost the S/N$_{\mathrm{int}}$ of all detections but this will not be compensated for by an increase in fainter sources. The result of this is that we cannot simply compare the \snint{} values of different surveys directly, but must transform the results to a common standard of sensitivity and spectral resolution.

\subsection{Correcting the \snint{} distributions}
\label{sec:modifysnint}
Fortunately it is possible to account for these considerations, at least to some degree. If we assume that the sources detected in a survey always have reasonably accurate measurements for their fluxes and velocity widths, we can recalculate their \snint{} to the value which would be found if detected in any other survey. I emphasise that this exercise is rather simplistic and purely for the sake of a fair numerical comparison. It does not constitute a rigorous appraisal of the overall performance of each survey -- for that, readers should consult the individual survey papers.

\begin{figure*}[t]
	\begin{center}
		\includegraphics[width=180mm]{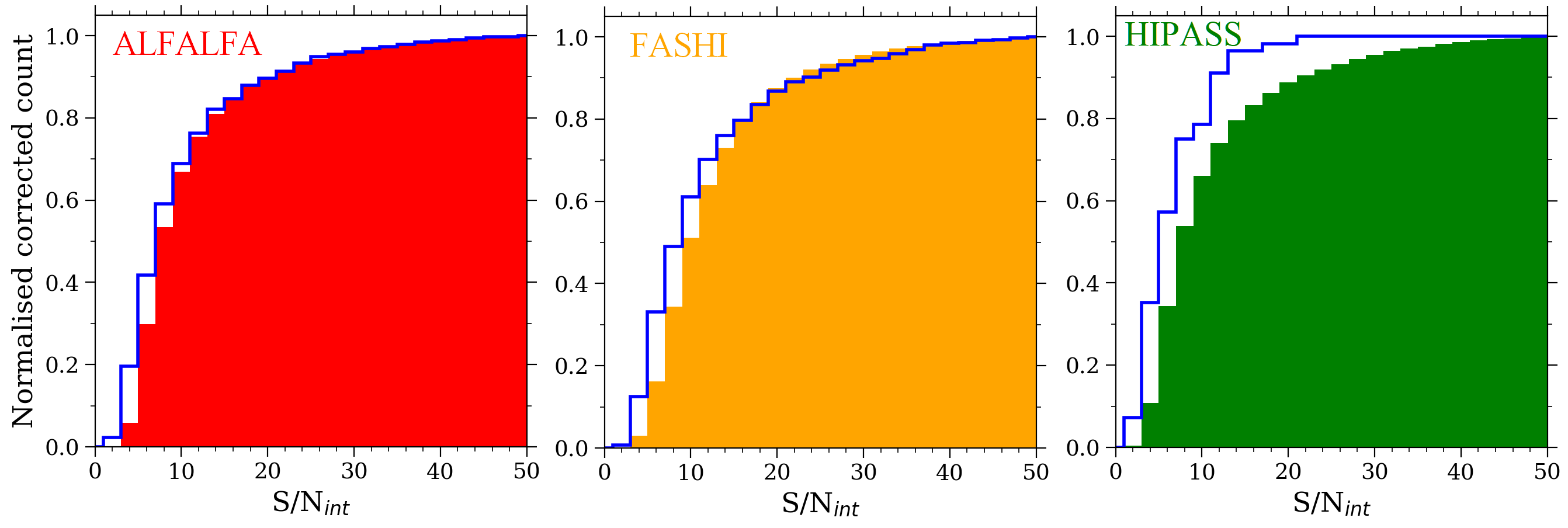}
		\caption[thing]{Cumulative, completeness-corrected integrated S/N distributions for selected surveys. AGES is the blue open line in all cases. From left to right : red is ALFALFA, orange is FASHI, and green is HIPASS. All curves have been corrected to degrade the calculated \snint{} values to a common standard, with the exact corrections described in section \ref{sec:modifysnint}.}
		\label{fig:allcorrect}
	\end{center}
\end{figure*}

Here, I recalculate the cumulative \snint{} distribution for each survey and compare with AGES. This is shown in figure \ref{fig:allcorrect}. The S/N$_{\mathrm{int}}$ range is limited to (somewhat arbitrarily) between 0 and 50, which removes the small-number tail of much brighter sources. Since the surveys all differ considerably, each comparison with AGES is plotted in a different panel. In each case, the S/N$_{\mathrm{int}}$ is recalculated by degrading the surveys to the worst parameter of those in each comparison pair. Specifically :
\begin{itemize}
	\item AGES-ALFALFA : the S/N$_{\mathrm{int}}$ of the AGES sources is recalculated using the AGES fluxes but at the ALFALFA \textit{rms}, with no corrections made to the ALFALFA measurements. Velocity resolutions of both surveys are identical.
	\item AGES-FASHI : here the AGES S/N$_{\mathrm{int}}$ uses the AGES fluxes and velocity resolution but the FASHI \textit{rms}, with the S/N$_{\mathrm{int}}$ for FASHI calculated using their own flux measurements and \textit{rms} levels but at the AGES velocity resolution.
	\item AGES-HIPASS : HIPASS S/N$_{\mathrm{int}}$ is calculated using the HIPASS measurements, while the AGES data is degraded to use the HIPASS \textit{rms} and velocity resolution but assuming the AGES flux values.
\end{itemize}
Two additional points are important when considering this figure. First, the histogram shows the normalised cumulative count as the footprint areas for the surveys differ substantially. For ALFALFA, which has covered the entire of the AGES footprint, I also tried comparing only those sources found in the AGES areas without normalising the counts. This approach, which ensures that AGES and ALFALFA were observing the same true source population (ground truth), gave virtually identical results to the figure shown here, demonstrating that this is an effective way to control for the different survey areas. Secondly, given the results of the fiducial experiment of section \ref{sec:results}, the completeness values of the blue error points in figure \ref{fig:c-fsn} are given in table \ref{tab:cc}. The AGES histograms account for this by multiplying the count in each \snint{} bin with the corresponding completeness values given in the table. In short, the aim of this figure is to compare the \snint{} for each survey with AGES, where we believe we have a good understanding of the completeness curve, after accounting for the survey parameters of sensitivity, resolution, and coverage area. In principle, remaining differences between the curves can be ascribed to source extraction efficacy, though there are many caveats to this as will be discussed below. I note that since this is derived from the final published catalogues, it should account for \textit{all} aspects of the source extraction procedure, e.g. the initial input parameters for the algorithm to the use of supplementary data and visual vetting of the source lists.

\begin{table}[h]
\begin{center}
\caption[cctable]{Binned completeness levels as shown by the blue points in figure \ref{fig:c-fsn}.}
\label{tab:cc}
\begin{tabular}{c c c}
\hline
\multicolumn{1}{c}{S/N$_{\mathrm{int}}$} &
\multicolumn{1}{c}{Completeness} &
\multicolumn{1}{c}{1$\sigma$ error} \\
\multicolumn{1}{c}{[bin centre]} &
\multicolumn{1}{c}{[\%]} &
\multicolumn{1}{c}{}\\
\hline
1.5 & 2.0 & 2.0\\
2.5 & 12.0 & 5.3\\
3.5 & 33.5 & 11.7\\
4.5 & 68.5 & 14.4\\
5.5 & 85.0 & 6.8\\
6.5 & 98.0 & 3.6\\
7.5 & 100.0 & 0.7\\
8.5 & 100.0 & 0.0\\
\hline
\end{tabular}
\tablefoot{The completeness levels are the median of each bin. All bins are of width 1.0.}
\end{center}
\end{table}

First, the difference between the raw values shown in figure \ref{fig:sndistro} and the corrected distributions in figure \ref{fig:allcorrect} is striking. In all cases, there is a much closer agreement between the corrected \snint{} distributions, demonstrating both that the raw \snint{} parameter cannot be used for direct survey-survey comparisons but also that appropriate corrections are possible. Figure \ref{fig:allcorrect} shows that, in all cases, the visual-based AGES detects a higher fraction of its sources at lower S/N$_{\mathrm{int}}$ values, consistent with the previous findings that visual extraction reaches higher completeness values for fainter sources.

Importantly, the closer agreement between the corrected distributions indicates that poor source extraction efficacy is not responsible for the much larger differences seen in the raw \snint{} distributions. Notably, \cite{fasthi} report that only 6\% of their sources have a native S/N$_{\mathrm{int}}$\,$<$\,10, which might suggest that they (either through their choice of parameters for SoFiA and/or their visual inspection of the initial catalogues) are doing significantly worse at recovering faint sources. However the corrected completeness curve shows that a significant contributing factor is FASHI's better velocity resolution, and in fact their source extraction is likely about as good as is currently possible. In the corrected completeness curve, FASHI finds about 30\% of its sources with S/N$_{\mathrm{int}}$\,$<$\,10 (far better than the uncorrected distribution), compared to AGES 50\%.

For both ALFALFA and FASHI the difference in the distributions compared to AGES is, after applying the corrections, slight. Conversely, there is still a substantial difference compared to the visual extraction favoured by AGES and the \textsc{multifind} extractor used by HIPASS, but this is not surprising. In an early test, \cite{multifind} found that \textsc{multifind} only recovered 46\% of the known real sources found by visual extraction. Additionally, HIPASS did not cross-check its candidate \HI{} detections with optical data during its initial source extraction (the larger beam size making the identification of a single optical counterpart more challenging), with this only done during a later analysis phase (\citealt{hipassoptical}). This could mean that the `recording bias' discussed in section \ref{sec:noticing} may have meant they were compelled to reject fainter sources for the sake of reliability, more than the extractor not detecting these objects at all.

Interestingly, \cite{multifind} note that higher noise levels can actually have a beneficial effect on detectability due to the `noise boosting' discussed in section \ref{sec:comprelvar}, with, counter-intuitively, the extractor better able to recover fainter sources in the \textit{noisier} data. In this case the extractor was using a peak flux threshold for detection rather than S/N, so while fainter sources would still likely have lower S/N levels in the noisier data, they would potentially have higher peak fluxes. Such an effect well illustrates how subtle source extraction techniques can be, emphasising that the specific details of the methodology can have surprising effects and must be accounted for in any comparative analysis : clearly it should not be taken as a general statement that noise enhances the detectability of faint sources, only that this is possible in some particular circumstances. \cite{multifind} found that the effect was strongest for wider sources, but the disadvantage is that what appears as a coherent structure to the eye is not necessarily interpreted as such by an algorithm. \cite{SOFIA2} note that SoFiA `fragments' as many as 40\% of sources of widths greater than 400\,\kms{}, while \textsc{multifind}'s completeness has a strong dependence on the spectral profile shape. 

The overall result of figure \ref{fig:allcorrect} is clear : visual extraction results in a greater fraction of its sources at lower \snint{} levels, once the recovered signals have been properly corrected to allow a fair comparison. This broadly supports the results of the direct tests which showed that the high reliability levels desired for automatic algorithms (at least those which were tested) can only be obtained at lower completeness levels than are reached by visual inspection. There are however many caveats to this result, which I discuss below.

\subsection{Other real-world considerations and caveats}
In section \ref{sec:biasproof} I considered the various sources of bias that may have affected the results of the visual extraction experiment. Similarly, while the quantitative value of the measured \snint{} can control for the differences in the surveys considered here, it is important to account for other differences that may affect the validity of the comparison.

\subsubsection{Differences in \snint{} not resulting from source extraction methodology}
While the survey-corrected \snint{} distributions show a much closer match than the raw values, the remaining differences may not all be due to source extraction methodology. Being of similar natures and at the same telescope, ALFALFA and AGES should have broadly similar data characteristics (especially in terms of RFI) to each other, but FASHI and HIPASS may well have qualitatively different data properties, both in comparison to each other and the Arecibo surveys. For example, as mentioned previously, \textit{rms} does not measure the coherency of structures in the noise; the shape of the spectral baseline may be different; RFI will almost certainly differ significantly at different sites; HIPASS covered the Galactic plane where the noise is significantly higher. While the majority of survey data should have comparable, uniform, random-Gaussian noise with reasonably flat spectral baselines, this is by no means true everywhere. Thus, while it is interesting that the corrected \snint{} distributions \textit{all} show a preference for faint sources in the visually-extracted catalogue, not all of this difference may result from the eye's natural ability to better discern real faint sources from spurious ones : conceivably the eye might well do worse in some situations. In combination with the direct comparisons of visual and automatic extractors, however, the results here add supporting, though not conclusive, evidence which is consistent with those findings.

\subsubsection{Independent ALFALFA completeness estimates}
\label{sec:aaowncomp}
\cite{AA40} perform their own completeness estimate on the 40\% ALFALFA catalogue. Their equations 4 and 5 give the estimated 90, 50 and 25\% completeness levels based on a combination of the flux and velocity width of the detections. Although they do not do so, these results can be transformed into \snint{} via equation \ref{eqt:aasn}. Since they adopt in their analysis a different break point in sensitivity with regard to velocity (log(W50)\,=\,2.5 or 316\,\kms{}) compared to that used in the \snint{} equation (400\,\kms{}), the completeness levels are not at a constant value of \snint{} -- however the variation is slight. The \snint{} levels are :

\begin{itemize}
\item 90\% completeness : \snint{}\,=7.04 for W50\,$<$\,316\,\kms{}; \snint{}\,=7.92 for W50\,$<$\,600\,\kms{}
\item 50\% completeness : \snint{}\,=5.91 for W50\,$<$\,316\,\kms{}; \snint{}\,=6.74 for W50\,$<$\,600\,\kms{}
\item 25\% completeness : \snint{}\,=5.57 for W50\,$<$\,316\,\kms{}; \snint{}\,=6.26 for W50\,$<$\,600\,\kms{}
 for W50\,$<$\,600\,\kms{}
\end{itemize}

From the visual experiments here, the completeness levels of 90, 50 and 25\% are found at \snint{} levels of 5.53, 3.93, and 3.25 respectively. Clearly the ALFALFA curve, though (as with SoFiA) converging at levels above 6--7, is much steeper than our visual extraction. This makes the difference in the \textit{raw} \snint{} distributions surprising, since ALFALFA find a greater fraction of fainter sources than the other surveys. Recall, however, the tests of section \ref{sec:noticetests}, which demonstrated that we would indeed find more faint sources in our data if they existed, and the \textit{corrected} \snint{} distribution in figure \ref{fig:allcorrect}, which shows -- in effect -- that an even higher fraction of faint sources could be found in ALFALFA by visual inspection. 

It is nonetheless unclear why AGES does not find such a high fraction of faint sources at its native \snint{} values. AGES has an \textit{rms} a factor 3.8 lower than ALFALFA, but the median flux level of its sources is only a factor 2.1 lower. To detect the same proportion of sources below a \snint{} of 9 (well above the completeness limit) would imply we have missed around 700 sources, which is scarcely credible. One possible explanation could be that our background volume, which gives the bulk of our detections, could be dominated by clusters where gas loss has truncated the \HI{} mass function. This requires a much more detailed investigation, and should await publication of the complete AGES Volume.

\subsubsection{Sensitivity of algorithms}
Intrinsic to the operation of both \textsc{multifind} and SoFiA is their ability to smooth the input data to increase source detectability. While they take as input cubes with the typical characteristics as given in table \ref{tab:surveytable}, smoothing means that the cube they \textit{actually search} will have a different sensitivity level. While smoothing will normally result in a decrease in \textit{rms} such that the \snint{} of a source will not change, both SoFiA and \textsc{multifind} ultimately use peak S/N for their detection thresholds, which \textit{does} change with smoothing.

This makes the comparison presented here -- arguably -- a best-case scenario for the automatic extractors. In a sense, for a strict comparison to visual extraction, either the automatic extractors ought to be run with smoothing disabled, or the visual search run on the smoothed data sets as well : otherwise, they are not really searching the same data. Disabling smoothing would likely result in a loss of detections of the fainter sources for the algorithms, shifting the \snint{} distributions to the right even though this parameter is not itself affected by smoothing. On the other hand the algorithms were designed from the outset to use smoothing, so disabling it would not be running them as intended and thus not presenting a fair comparison. The difficulty in deciding the best approach to compare the results is that visual and algorithmic source extraction work by fundamentally different methods. By-eye searches examine whole images in a way which is still not fully understood (but see section \ref{sec:sum}), whereas algorithms rely heavily on individual data values albeit after applying different smoothing levels. Whether the eye itself already applies a form of smoothing similar to that used by algorithmic searches is an open question.

\subsubsection{AGES in anger : the importance of training}
\label{sec:angryages}
The fiducial experiment was only carried out by the author, and an obvious follow-up study would be to investigate how the completeness rate varies between individuals both with and without training. While I cannot here fully address those points, fortunately there is some data from AGES which gives a clue as to their importance.

In total, 16 different people have been involved to varying degrees with AGES source extraction, and in some cases we have detailed records of their different completeness rates. In \cite{agesiv}, three people inspected the preliminary NGC 7448 data set (using \textit{kvis}), finding between 141 and 176 sources each (for a grand total of 191 unique sources). This is a rather large variation, with one person missing 26\% of the visually detected sources. However since the advent of FRELLED, and implementing a standard training scheme, this has been drastically reduced. New observers are first trained on the \cite{agesvc2} (VC2) data set, which was previously searched with multiple techniques and all uncertain sources found therein were subjected to follow-up observations. Everyone therefore gets experience in finding real sources at different S/N levels, in determining whether their candidates are real or likely spurious, and in accounting for the varying noise levels caused by RFI in certain parts of the data cube.

In comparison with \cite{agesiv}, the detection rates of the observers in \cite{agesvii} (which introduced FRELLED) were much more similar. Here two observers searched the cube independently and each found 18 or 20 sources (5\% of the final total of 334) that the other missed. For the WAVES data set (Part\'{i}k et al. in preparation, a larger area than the preliminary observations reported in \citealt{waves}), one observer made an initial catalogue and masked the sources, with another then searching this masked data set. The second observer found 15 additional candidates but after discussions it was agreed that only one of these (out of 55 other sources) was at all likely to be real, based on its optical counterpart rather than the quality of the \HI{} signal. Similarly in the complete Abell 1367 data (\citealt{boris}), the second observer searched the masked cube and discovered only 10 more candidate sources (2.5\% of the 412 visually-identified objects). Finally, one observer managed to detect a single new source in the VC2 training data, out of a total of 49 objects. 

While the importance of training is discussed in, for example \cite{sports}, there have been few formal studies of its efficacy or the intrinsic variation between observers. Overall, from the limited data here, it does not appear that there are great variations between individuals when it comes to visually identifying sources, so long as proper data inspection techniques and standardised training methods are used.

\section{Summary and discussion}
\label{sec:sum}
This work describes the results of an experiment to quantify the completeness and reliability of visual source extraction, in the context of extragalactic \HI{} surveys. A sample of 8,500 galaxy-free data cubes were injected with a total of 4,232 sources. This explored the parameter space of line width (from 25 to 350\,\kms{} in steps of 25\,\kms{}) and S/N (from 1.5 to 6.0 in steps of 0.25). For each pair of S/N and line width values, 100 galaxy-free data cubes were extracted at random from a larger master cube, and each was injected with either one or zero sources of the specified parameters. I performed a visual search to recover these sources, the goal being to rigorously and precisely quantify the efficacy of visual source extraction.

The results of this search largely support the findings of studies using real observational data. Previous ALFALFA and AGES studies have found a distinct threshold in terms of flux and line width, above which detections are much more probable to be genuine astrophysical sources than spurious identifications. This has largely been ascertained empirically, using corroborating multi-wavelength data in conjunction with follow-up \HI{} observations. Here this has been examined using the extremely controlled conditions of precisely-known artificial sources. While many automatic extractors have previously been tested in this way, little similar work has been done to determine how applicable this finding is with regards to visual source extraction. We have compared visual and automatic techniques in relative terms, but made few assessments of the \textit{absolute} capabilities of visual extraction by itself.

I find that the integrated S/N parameter of S07 is an excellent way to characterise the results of visual source extraction. There is a clear, tight correlation between the integrated S/N and completeness, regardless of either peak S/N or line width. Numerous tests were performed to control for the biases in the main experiment, including eliminating any foreknowledge of the probability a source was injected, randomising source properties, injecting sources of different detectability into a single cube, and injecting them into cubes with real galaxies present. All gave results in good agreement with the main experiment, demonstrating that the effects of these biases are minor. Far from being subject to unknown variations, visual extraction is a predictable, measurable process which gives statistically robust, objective results. 

Given the increased difficulty of obtaining follow-up observations of Arecibo discoveries, which are still being uncovered from archival data sets, using parameters like this to assess completeness and reliability will become increasingly unavoidable. The S/N$_{\mathrm{int}}$ parameter can also be used to provide reassurance that other visually-extracted catalogues in older papers are reliable even when follow-up observations are lacking, or the original data unavailable for reanalysis. 

Of course there is some scatter in the completeness rates as a function of S/N$_{\mathrm{int}}$, but these variations are small, and this is not a problem to which automatic extractors are wholly immune. \cite{SOFIA2} stress that SoFiA's performance is strongly dependent on the data quality, with one data set returning results at essentially 100\% reliability and another at `only' 75\% (though this is still far above the automatic extractors assessed in \citealt{agesvc2}). Both \cite{MeerGPS} and \cite{fasthi} also note that they visually inspect SoFiA's output to ensure reliable results. Importantly, \cite{wallabyCNN2} note that their convolutional neural network, which they use for vetting SoFiA's output, `can only approach, but not surpass, the accuracy of human experts.'

Overall, the completeness and reliability rates of visual extraction compare very favourably to the automatic methods directly tested here (GLADoS and SoFiA). GLADoS is the weakest of the three, being faster than visual extraction, but less reliable and complete. SoFiA is very much faster than GLADoS and achieves substantially better completeness and reliability. Above a \snint{} level of approximately 6.5, SoFiA and visual extraction both reach completeness levels approaching 100\%. Below this, visual extraction outperforms SoFiA in terms of reliability, but the greater speed of SoFiA may be sufficient compensation to make it a valuable alternative : the longer candidate list generated by SoFiA for these fainter sources is nonetheless easier and faster to inspect than a raw data cube.

This result is supported by inspection of the S/N$_{\mathrm{int}}$ distributions from various surveys. Importantly, the distributions of the raw values of \snint{} cannot be directly compared between different surveys, but a meaningful comparison can be made between survey pairs once corrected for the different observational parameters. Once these corrections are applied, visual extraction tends to reach higher completeness values at lower S/N levels than automatic techniques. A probable explanation for this is the difficulties for algorithmically measuring faint, wider sources, a problem also noted for \textsc{multifind} and SoFiA. While automatic searches have greatly improved in recent years and can now compete with visual extraction for completeness, it still requires significant innovation for an algorithm to actually \textit{improve} on visual results. Their principle advantage is cataloguing speed (at which they excel compared to humans), not raw detection capability.

Beyond astronomy, the mechanism by which visual pattern recognition works despite the presence of noise has been explored by psychologists. Further to the point that visual extraction is not unreliable, \cite{scanpaths} show that even the movement of the eye can be predicted when asked to retrieve information. A study by \cite{camo} has even more direct relevance. When training subjects to break camouflage (in which signals are deliberately concealed by noise), they found that human pattern recognition essentially learns the statistical properties of the background, allowing the eye to detect any deviation from the established trend. This may help explain why humans are generally good at discriminating signals even in the presence of strong RFI. \cite{camo} also demonstrate the importance of training, with subjects showing measurable and consistent improvements in breaking camouflage as a result of inspecting more training data prior to the search.

The supposed inconsistency of visual extraction can be controlled through proper training and cataloguing tools. When we search cubes fully independently the catalogues differ at the level of only a few percent, and when a second observer searches an already-masked cube, they typically find an even smaller fraction of additional detections. The ability to mask sources is crucial for two reasons. First, it prevents each observer from becoming confused as to which sources they have already catalogued. Second, it ensures that with each successive observer completeness can only ever increase, thus mitigating the effect of variation in individual source extraction skills. The practical result is that most people do about equally well at visual extraction, when they have this masking capability available.
 
Moreover, the experiment presented here points to visual extraction as a rapid process. As discussed in \cite{FRELLED5}, the time to identify a source by eye is not the limiting factor at all. Rather it is the speed with which the visual identification can be converted into a record which is the bottleneck. In this setup, with the operations needed to catalogue sources reduced to a minimum, each batch of 100 cubes took approximately 30 minutes to catalogue, i.e. about 800 sources catalogued per working day. Extrapolating, this would mean ten people could catalogue an ALFALFA-sized data set inside a week. This is clearly optimistic, with real-world factors such as visual fatigue and source density obviously able to strongly influence cataloguing speed. Nonetheless, while the importance of rapid automatic algorithms such as SoFiA cannot be doubted, this result indicates that by-eye searches may yet play a role even in an era increasingly dominated by larger data sets.

\section*{Data availability}
The source code and galaxy-free Mock spectrometer data used in this work are available at the following URL : \url{https://tinyurl.com/yzv6b5f2} (80 MB). This includes instructions for running the code for anyone wishing to recreate the experiment for themselves, both the fiducial experiment and the supplementary tests (in particular, `supplementary experiment 3' may be useful as a training tool for new observers). The complete 200 GB of processed cubes and FRELLED files used here can be made available on request to the author. 

\begin{acknowledgements} 
I thank the anonymous reviewer whose suggestions substantially enhanced the quality of the manuscript. I also thank my generous colleagues who read the article for their constructive (and occasionally facetious) feedback, which has probably improved the manuscript. Special thanks to Robert Minchin for productive discussions which cleared up several misconceptions.

This work was supported by the institutional project RVO:67985815 and by the Czech Ministry of Education, Youth and Sports from the large Infrastructures for Research, Experimental Development and Innovations project LM 2015067. 

This work has made use of the SDSS. Funding for the SDSS and SDSS-II has been provided by the Alfred P. Sloan Foundation, the Participating Institutions, the National Science Foundation, the U.S. Department of Energy, the National Aeronautics and Space Administration, the Japanese Monbukagakusho, the Max Planck Society, and the Higher Education Funding Council for England. The SDSS Web Site is http://www.sdss.org/.

The SDSS is managed by the Astrophysical Research Consortium for the Participating Institutions. The Participating Institutions are the American Museum of Natural History, Astrophysical Institute Potsdam, University of Basel, University of Cambridge, Case Western Reserve University, University of Chicago, Drexel University, Fermilab, the Institute for Advanced Study, the Japan Participation Group, Johns Hopkins University, the Joint Institute for Nuclear Astrophysics, the Kavli Institute for Particle Astrophysics and Cosmology, the Korean Scientist Group, the Chinese Academy of Sciences (LAMOST), Los Alamos National Laboratory, the Max-Planck-Institute for Astronomy (MPIA), the Max-Planck-Institute for Astrophysics (MPA), New Mexico State University, Ohio State University, University of Pittsburgh, University of Portsmouth, Princeton University, the United States Naval Observatory, and the University of Washington.

\end{acknowledgements}

\bibliographystyle{aa}
\bibliography{references}

\end{document}